\documentclass{article}
\usepackage{moreverb}
\usepackage{array}
\usepackage{tikz}
\usepackage[numbers]{natbib}
\usepackage{booktabs}
\usepackage{listings}
\usepackage{amsmath}
\usepackage{amsfonts}
\usepackage{algorithm}
\usetikzlibrary{arrows.meta}
\usepackage{graphicx} %% for Jpegs\usetikzlibrary{arrows.meta}s
\usetikzlibrary{shapes,positioning}
 \usepackage{url}
\usepackage{moreverb}
\newcommand\BibTeX{{\rmfamily B\kern-.05em \textsc{i\kern-.025em b}\kern-.08em
T\kern-.1667em\lower.7ex\hbox{E}\kern-.125emX}}

\usepackage{subcaption}

\usepackage{enumitem}

\usepackage{algpseudocode,algorithm}

\makeatletter
\algrenewcommand\ALG@beginalgorithmic{\normalfont}
\makeatother

\begin{document}

\title{Comparing methods to assess treatment effect heterogeneity in general parametric regression models}

\date{}
\author{Yao Chen\textsuperscript{*}\thanks{Advanced Methodology and Data Science, Novartis Pharmaceuticals Corporation, East Hanover, New Jersey, USA}
\and Sophie Sun\textsuperscript{*}\thanks{Advanced Methodology and Data Science, Novartis Pharmaceuticals Corporation, Cambridge, MA, USA}
\and Konstantinos Sechidis\thanks{Advanced Methodology and Data Science, Novartis Pharma AG, Basel, Switzerland}
\and Cong Zhang\thanks{China Novartis Institutes for Bio-Medical Research Co, Shanghai, China}
\and Torsten Hothorn\thanks{Biostatistik und Prävention, Universität Zürich}
\and Björn Bornkamp\footnotemark[3]
}
\maketitle

\abstract{This paper reviews and compares methods to assess treatment effect heterogeneity in the context of parametric regression models. These methods include the standard likelihood ratio tests, bootstrap likelihood ratio tests, and Goeman's global test motivated by testing whether the random effect variance is zero. We place particular emphasis on tests based on the score-residual of the treatment effect and explore different variants of tests in this class. All approaches are compared in a simulation study, and the approach based on residual scores is illustrated in a clinical trial with time-to-event outcome comparing treatment versus placebo. Our findings demonstrate that score-residual based methods provide practical, flexible and reliable tools for exploring treatment effect heterogeneity and treatment effect modifiers, and can provide useful guidance for decision making around treatment effect heterogeneity.}


\maketitle

\renewcommand\thefootnote{}
\footnotetext{*These authors contributed equally to this work.}

\renewcommand\thefootnote{\fnsymbol{footnote}}
\setcounter{footnote}{1}

\section{Introduction}
\label{sec:intro}

Assessment of treatment effect heterogeneity (TEH) across patient subpopulations in RCT (Randomized Clinical Trials) is an important, but challenging task for clinical trial sponsors. Typically the clinical trial design, sample size and analysis strategy are not explicitly planned to assess the treatment effect within subgroups. As a result, generally, the data from the RCT provide limited information on treatment effects in subgroups due to the limited sample size. In addition investigating multiple subgroups causes a severe multiplicity problem. These factors may explain why subgroup findings in clinical trials often fail to be replicated (if replication was tried) \cite{Yusuf1991, wallach2017evaluation}.

Another important consideration when assessing treatment effect heterogeneity is that the presence and magnitude of TEH depend heavily on the chosen effect measure. In Table \ref{tab:scale_example} we show a numerical example where the subgroup has a larger treatment effect than the complementary group on the absolute difference scale, but on the relative risk scale the complementary group has a larger treatment effect. On the odds-ratio scale finally both subgroup and complement group have the same treatment effect. This scale dependency is especially relevant in cases of quantitative interaction, where all subgroups benefit in the same direction but to different extents. Therefore one needs to consider which treatment effect scale should be chosen for the investigation of TEH. 

\begin{table}
    \centering
    \begin{tabular}{|c|cc|ccc|} \hline
    &   \multicolumn{2}{c|}{Response rates} &
      \multicolumn{3}{c|}{Effect measure} \\
    & New drug & Control & difference & ratio & odds-ratio \\ \hline
        Subgroup   & 0.8 & 0.333 & 0.467 & 2.4 & 8 \\
        Complement & 0.25 & 0.04 & 0.21 & 6.25 & 8 \\ \hline
    \end{tabular}
    \caption{Numerical example illustrating that the magnitude of treatment effect depends on the treatment effect measure (from FDA covariate adjustment guidance\cite{FDA:2023}). The considered outcome here is a binary outcome summarized with response rates per treatment and subgroup.}
    \label{tab:scale_example}
\end{table}

Several considerations may play a role when choosing an effect measure. Recently, some emphasis has been placed on targeting the conditional average treatment effect (CATE) \cite{lipkovich2024modern, sechidis2025using}, which is given by the difference in conditional expectations of the outcome (conditioning on specific covariate values). This effect measure has the advantage of being defined without reference to a specific model. From a public health perspective it has been argued that this difference (absolute) scale is the most relevant scale to measure treatment effects\cite{vanderweele2014tutorial}. The EMA guideline on the investigation of subgroups in confirmatory clinical trials\cite{ema:2019} suggests to use the treatment effect scale based on the models which are commonly used and to consider the scale where the treatment effect is more consistent. On the other hand it also suggests to consider alternative scales in particular for benefit/risk evaluations. The use of regression models, such as for example logistic regression or the Cox model, is very wide-spread as primary analysis method in clinical trials, with subgroup treatment effects (for example in forest plots) being produced on the model-
based treatment effect scale implied by the primary analysis model. We hence believe that it is of value to investigate methods for assessing heterogeneity for general regression model-based effect measures.

\subsection{WATCH workflow}
Various workflows have been proposed to address the underlying issues from different perspectives\citep{rube:shen:2015, lipk:dmit:agos:2017, muysers2020systematic, kent2020a}. In this paper, we would like to specifically focus on the Workflow for Assessing Treatment effeCt Heterogeneity (WATCH)\cite{sechidis2024watch} and compare different statistical methods on how to implement this workflow in the context of treatment effect measures defined by regression models. 

\begin{figure}
    \centering
    \includegraphics[width=0.33\linewidth]{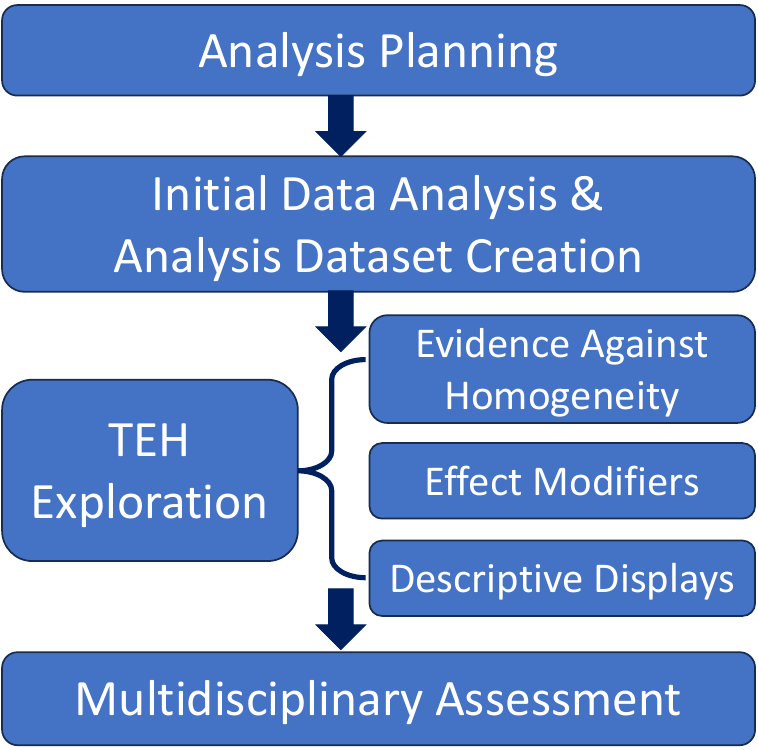}
    \caption{Overview of WATCH workflow and the four main steps.}
    \label{fig:watch}
\end{figure}

WATCH focuses on \textit{exploratory} assessments of treatment effect heterogeneity in clinical drug development and consists of four steps (see Figure \ref{fig:watch}): 1) Analysis Planning, 2) Initial Data Analysis and Analysis Dataset Creation, 3) TEH Exploration, and 4) Multidisciplinary Assessment. The workflow addresses the challenges of limited sample size and multiplicity by stressing the importance of adequate pre-planning, incorporating a-priori, external information, and ensuring appropriate communication. For example, it is strongly suggested to document the external evidence of potential treatment effect modifiers before data-base lock of the underlying RCT or at least before start of the data analysis. Additionally, the importance of multidisciplinary discussions is emphasized. Data-based findings that cannot be explained by previous external evidence and/or data, will be considered more speculative than those where external evidence is available. Furthermore the approach emphasizes the need for appropriate communication, for example by avoiding strong confirmatory statements for treatment effects in subgroups that are not appropriate for this exploratory investigation. Instead, focus is on assessing the overall evidence against homogeneity and identification of potential effect modifiers. The aim is to provide a global overview rather than focusing on specific findings.

In this paper we discuss implementation of this core data-analytical step 3, TEH Exploration, which consists of three further sub-steps: 

\begin{enumerate}
    \item[(a)] \textit{Evidence Against Homogeneity}\\ This step provides an overall assessment of whether there is systematic variation in treatment effects across defined patient baseline covariates (\textit{i.e.} more systematic than expected under a homogeneous model). A global heterogeneity test implicitly accounts for the multiplicity of the included variables and helps determine the strength of evidence against homogeneity by reporting a $p$-value against the hypothesis of a homogeneous treatment effect. We suggest to interpret the $p$-value on a continuous scale, such as the surprise scale\citep{cole2020}, rather than a binary decision rule.
    \item[(b)] \textit{Effect Modifiers}\\ Describing which patient baseline variables are associated with the treatment effect is important for understanding how the treatment effect varies across different patient subgroups in the observed data. The output of this step is a variable importance ranking. 
    \item[(c)] \textit{Descriptive Displays}\\ Creating graphical displays to describe treatment effect heterogeneity helps to visualize the variation in treatment effects across different baseline covariates. These visualizations are crucial for interpreting the data and communicating findings in an understandable manner. Of course the result from  step (a) should be considered in interpreting the results from (b) and (c) In case of low evidence against homogeneity, the variable importance ranking and also observed treatment effects are more likely driven primarily by noise and the corresponding results should be communicated having this in mind..
\end{enumerate}

\subsection{Our contribution}
Regression-based effect measures remain widely used in clinical trials, particularly for binary, count, and time-to-event outcomes. Therefore, it is crucial to investigate methods for assessing heterogeneity for general regression model-based effect measures. In this paper we will consider which statistical methods to use to implement step (a) and (b) in the context of treatment effect measures from general parametric regression models, for continuous, binary, count and time-to-event outcomes. Recently different methods for implementation of steps (a) and (b) were compared\cite{sun2024comparing}. However, the authors only considered continuous and binary outcomes and in the conditional average treatment effect scale (CATE). In our work, we extend the scenarios to count and time-to-event outcome and consider the model-based effect measures. We also propose to use score residual based approach to achieve step (a) and (b) and evaluate its performance on both simulation and real case. 

The outline of this paper is as follows: Section \ref{sec:methodology} discusses the interpretation of regression model-based treatment effect measures, global tests for TEH in regression models and reviews different methods for investigating model-based treatment effect heterogeneity, with a focus on score residual based methods. Section \ref{sec:simulations} compares different variants of the score residual approach to other methods for assessing steps 3a) global heterogeneity test and 3b) variable importance for treatment effect modifiers in a simulation study. Section \ref{sec:example} illustrates the score residual method in a example with time-to-event outcome. Section \ref{sec:concl} concludes.

\section{Methodology}
\label{sec:methodology}

In this section, we first clarify the treatment effect estimand underlying regression-based treatment effect measures and review global interaction (global TEH) tests. We specifically evaluate an approach for characterizing deviations from treatment effect homogeneity based on score residuals, which is closely connected to split criteria used in model-based partitioning trees \cite{zeil:hoth:horn:2008, seib:zeil:hoth:2016}. While the main aim of these works was to build an overall outcome model, our work has a different objective. We aim to derive (i) a global test for TEH and (ii) assess variable importance for TEH. 

\subsection{Notation}

Assume we observe data $(y_i,z_i,\mathbf{x}_i)$ for $\; i \in \{1,...,N\}$ where $y_i$ is the outcome of interest, $z_i\in \{0,1\}$ is a binary indicator for treatment and $\mathbf{x} \in \mathbb{R}^k$ is a vector of baseline covariates.  Consider that we fit a model $m$ to minimize the loss function $$\sum_{i=1}^N \Psi(\boldsymbol{\beta}_{\mathbf{x}}, \delta, \xi\mid y_i, \mathbf{x}_i)$$ where $\boldsymbol{\beta}_{\mathbf{x}}$ is the coefficient for baseline covariates, $\delta$ is the overall treatment effect, and $\xi$ is the nuisance parameter. Most commonly used models in clinical trials are of this form. In a generalized linear model (GLM) the loss function $\Psi$ is the negative log-likelihood function and the model equation for patient $i$ is 
$E(y_i\mid z_i,\mathbf{x}_i) = g(\eta_i) = g(\beta_0+\mathbf{x}_i'\boldsymbol{\beta}_\mathbf{x}+\delta z_i)$, with $g(.)$ being a inverse link function, $\beta_0$ being the intercept (corresponding to $\xi$ in the loss function), and $\eta_i$ being the linear predictor. In a Cox model, the model is $\log(h_i(t)) = \log(h_0(t)) +\mathbf{x}_i'\boldsymbol{\beta}_\mathbf{x}+\delta z_i$, with $h_i(.)$ being the hazard function for patient $i$ and $h_0(t)$ the baseline hazard function (corresponding to $\xi$). This is a nuisance parameter in the Cox model and the partial likelihood does not depend on it.

\subsection{Treatment effect measures for regression models}
\label{sec:treatm-effect-meas}

It is known that the maximum likelihood estimate for a generalized linear model $m$, under certain assumptions, such as independent identically distributed sampling, is consistent and asymptotically normally distributed for the pseudo-true parameter vector $(\boldsymbol{\beta}^*_{\mathbf{x}m}, \beta^*_{0m},\delta^*_m)$, which minimizes the Kullback-Leibler divergence of the fitted model $m$ with respect to the true data generating model $\mathcal{P}$\cite{fahrmeir1990maximum}. This implies that also the asymptotic targeted treatment effect estimand $\delta^*_m$ in most situations depends on specifics of the fitted model $m$, such as the parametric assumptions utilized, the covariates $\mathbf{x}$ and the form in which the covariates are adjusted for as main effects. 

An illustration for that is the hypothetical population distribution in Table \ref{tab:scale_example}. Assume that subgroup and complement each make up 50\% of the overall population. If a sample would be taken from this population and one would fit a logistic regression that does not adjust for subgroup as main effect, one would asymptotically target the pseudo-true parameter value of $\delta_m^*=\log(\frac{p_1/(1-p_1)}{p_0/(1-p_0)})\approx 1.57$, where $p_1=0.5\times 0.8+0.5\times 0.333$ and $p_0=0.5\times 0.25+0.5\times 0.04$. If one would adjust for the subgroup indicator as main effect, one would however target the value of $\delta_m^*=\log(8)\approx 2.08$ with the parameter of the treatment effect in the regression model. Both values are the true underlying values following from the true underlying data-generating model, but differ as different models are used for analysis (one adjusts for covariate, the other does not).

This example illustrates that the targeted estimand depends on the analysis model, for example the included covariates and specific type of adjustment and does not precede the model. This is an inherent consequence of the setting studied in this paper. Note that in the context of randomized trials it has been characterized, for which models, link functions and situations $\delta^*_m$ does not depend on the utilized additional covariates, or where the impact would be limited\cite{gail1984biased}. 

Note that also the notion of whether or not TEH exists depends on the utilized model, for example the covariates $\mathbf{x}$ included in model $m$. Given these limitations we propose to define the estimand as the coefficient $\delta_m$ in the regression model $m$ with $\eta_i=\beta_0+\mathbf{x}_i'\boldsymbol{\beta}_\mathbf{x}+\delta z_i$ (or $\log(h_i(t)) = \log(h_0(t)) +\mathbf{x}_i'\boldsymbol{\beta}_\mathbf{x}+\delta z_i$ for Cox model, resulting in the log hazard ratio), where $\mathbf{x}$ contains all potential effect modifiers included in the TEH analysis, based on a-priori, external evidence. This is justified, because in most randomized trials the stratification factors and main (=prognostic) effects that are adjusted for in the primary analysis model have been carefully selected a-priori (\textit{i.e.} data-independent). Furthermore, if the WATCH workflow is followed, a thoughtful selection of potential effect modifiers has been performed a-priori based on external evidence. 

\subsection{Global TEH tests for regression models}
\label{sec:global-teh-tests}

A global TEH test in context of a regression model can be performed by comparing the likelihood of the model $\eta_i=\beta_0+\mathbf{x}_i'\boldsymbol{\beta}_\mathbf{x}+\delta z_i$ from model $m$ with a model that utilizes

\begin{equation}
\label{eq:model2}
    \eta_i=\beta_0+\mathbf{x}_i'\boldsymbol{\beta}_\mathbf{x}+\delta z_i + z_i\mathbf{x}_i'\boldsymbol{\gamma}_\mathbf{x}.
\end{equation}

Compared to main effect model, model (\ref{eq:model2}) considers interaction effects between each covariate to the treatment where $\boldsymbol{\gamma}_\mathbf{x}$ is the interaction effect. It is well known that asymptotically an appropriately scaled difference of the log-likelihood function values of the two compared models are chi-squared distribution with $k$ degrees of freedom (see Section 4.5 in Davison (2003)\cite{davison2003statistical}). Unfortunately even for the case of moderate dimensional covariates considered in this paper, this approximation can be poor, as observed for binary data in realistic simulation settings\cite{sun2024comparing}. One way to approximate the finite sample distribution of the likelihood ratio statistic are parametric bootstrap tests (see Section 4.2 in Davison and Hinkley (1997)\cite{davison1997bootstrap}). Here the model $m$ is fitted, new data-sets are generated from this model and in each time model $m$ and \eqref{eq:model2} are fitted and their likelihoods are compared. This then provides an approximation of the null distribution.

An alternative approach is Goeman's global test\cite{goeman2004global}. It was specifically developed for high-dimensional prognostic covariates for outcome modeling, but can also be used in the context of global interaction tests \cite{callegaro2017testing}. It reformulates the test in the context of random effect models and assesses whether the common random effect variance for the interaction covariates is 0. 

In our simulations in Section \ref{sec:simulations} we will compare these three approaches. Note that all methods in this section are based on \eqref{eq:model2} and do not allow for higher order interactions, but are based on the additive (generalized) linear setting. 

\subsection{Using score residual to assess treatment effect heterogeneity}
\label{sec:using-score-residual}

When a fitted overall model $m$ is available, an alternative way to assess TEH is to calculate the derivative of the score function for each patient with respect to $\delta$, the treatment effect parameter. This is given by 
\begin{equation}
s_i=\left.\frac{\partial}{\partial \delta}\Psi(\widehat{\boldsymbol{\beta}}, \delta\mid y_i,\mathbf{x}_i )\right|_{\delta=\widehat{\delta}},
\end{equation}
where $\boldsymbol{\beta}=(\beta_0,\boldsymbol{\beta}_\mathbf{x}')'$ for generalized linear models and $\boldsymbol{\beta}=\boldsymbol{\beta}_\mathbf{x}$ for the Cox model.

We will use the term score residual for $s_i$, as it is commonly used in the survival analysis literature. By definition of the maximum likelihood estimate we have that $\sum_{i=1}^N s_i=0$, and the sign of the score residual $s_i$ indicates, for each patient, how a change in $\delta$ (from $\widehat{\delta}$) would change the likelihood contribution for this particular patient.

Note that the idea of using the score residuals (with respect to all model parameters) to assess and improve model fit has been used in papers by Hothorn, Zeileis and co-authors\cite{coin2008, zeileis2008model}. There the authors use independence tests or fluctuation tests as a splitting criterion within a tree-based partitioning approaches to select the variable to split in each node. The test that is performed at the root node can be interpreted as global test, whether there is any change in any of the model parameters (according to the score residuals). Assessing score residuals only with respect to the treatment effect parameter has also been mentioned in further papers \cite{seib:zeil:hoth:2016, thomas2018subgroup, sun2024comparing}, we investigate this approach a bit deeper in this paper.

The idea of using residuals to improve the model fit is also used in gradient boosting\cite{friedman2001greedy}. From that perspective the score residuals can be considered the gradient descent direction for improving the likelihood of the original model $m$ when allowing for individual treatment effects. Contrary to boosting only a single step is produced here.

An important point to consider is also the parameterization of the treatment effect. For that purpose consider the normal log-likelihood with $\sigma = 1$. Here we have that

\begin{equation}
    \label{eq:score_resid}
s_i=\left.\frac{\partial}{\partial \delta} \Psi(\widehat{\boldsymbol{\beta}}, \delta \mid y_i,\mathbf{x}_i)\right|_{\delta=\widehat{\delta}}=(y_i-\widehat{\beta}_0-\mathbf{x}_i'\widehat{\boldsymbol{\beta}}_\mathbf{x}-\widehat{\delta} {z}_i){z}_i = r_i{z}_i.
\end{equation}

Here $r_i=y_i-\widehat{\beta}_0-\mathbf{x}_i'\widehat{\boldsymbol{\beta}}_\mathbf{x}-\widehat{\delta} {z}_i$ is the least-squares residual. This implies that patients on control have $s_i=0$ (as $z_i = 0$). So the score residuals with respect to the treatment indicator contribute no information for patients on control. To remediate this we propose to use the centered treatment indicator $\Tilde{z}_i$ instead of $z_i$. The centered treatment indicator variable is defined as $\Tilde{z}_i=z_i-E(z_i\mid x_i)$\cite{dandl2024makes, gao2021estimating}. Note that in randomized trials $E(z_i\mid x_i)$ is known as the randomization probability distribution is known. This reparameterization of model $m$ does not lead to a change in the overall model fit (likelihood value), just the change of numerical value and interpretation of the intercept estimate: For the centered parameterization it represents an overall average outcome rather than the control arm outcome. This parameterization also makes the estimates of the intercept and the treatment effect less dependent (for the case of a linear homoscedastic regression even uncorrelated). This allows to single out the information on the treatment effect completely in the coefficient of the (centered) treatment indicator.

When using the centered treatment indicator $\Tilde{z}_i$ instead of $z_i$ in equation \eqref{eq:score_resid} the score residual $s_i$ also has a more intuitive interpretation: If $r_i>0$ it means that the model "underestimates" the observed outcome data for this patient. If the patient is on treatment (\textit{i.e.} $\Tilde{z}_i>0$ and thus also $s_i>0$) an increase in $\delta$ would improve the model fit: It looks as if the individual treatment effect for this patient is higher than predicted by the model. If the patient is on control $\Tilde{z}_i<0$ and thus the score residual $s_i<0$. So for this patient it seems the treatment effect is lower than predicted by the model (as their value on control is larger than predicted), so decrease in $\delta$ would be suggested to increase the individual likelihood. A similar reasoning holds for the case of $r_i<0$. In this way the score residuals $s_i$ contain information on the individual treatment effect for each patient. A positive score corresponds to a larger individual treatment effect and a negative score corresponds to a smaller individual treatment effect compared to the overall effect.

So, if the score residuals $s_i$ correlate systematically with certain baseline covariates, this hints that systematically for certain patients the overall treatment effect $\widehat{\delta}$ is too low (or too high). While if the score residuals fluctuate in a non-systematic way, there is no evidence against homogeneity, according to the covariates that are considered.

We can hence utilize similar approaches as proposed in Sechidis et al.(2025)\cite{sechidis2025using} for global testing for heterogeneity and for assessing variable importance. In that paper first a pseudo-observation of the individual treatment effect was derived for each patient (depending on their covariate values) and then global tests and assessment of variable importance was performed on the pseudo-observations of the treatment effect. Here the score residuals of the treatment effect $s_i$ will serve this same purpose.

Hence an independence test of the score residuals $s_i$ versus covariates allows for a global heterogeneity test of the treatment effect (\textit{i.e.} questions 3a from Section \ref{sec:intro}). One option to perform these tests are permutation tests. We consider the general framework reviewed in\cite{hothorn2006lego} and implemented in the \textbf{coin} \textsf{R} package\cite{coin2008}. In the default implementation linear statistics of the form 
\begin{equation}
    \boldsymbol{T}=\sum_{i=1}^N \boldsymbol{x}_i s_i\in \mathbb{R}^k
\end{equation}
are used. Permuting the covariates $\boldsymbol{x}_i$ versus the outcome $s_i$, one can derive the null distribution of the vector $\boldsymbol{T}$. To reduce this to a univariate statistic (and hence perform a global test), the \textbf{coin} package then uses either the maximum or the quadratic form of the standardized version of $\boldsymbol{T}$ as final test statistic. Note that asymptotic normal approximations for the permutation distribution of the test statistic are implemented in \textbf{coin} \textsf{R} package\cite{strasser1999}, which we will use in what follows.

Using a centered treatment effect indicator is a way to condense all information on the treatment effect in a single, more independent model parameter and make the parameter estimates for the treatment effect (and the score residuals across the parameter estimates) less dependent. An alternative approach to achieve this, is to initially orthogonalize the  scores\cite{zeileis2008model}, where fluctuation tests are then used. This approach can equally be used as a global test. We will also investigate this approach in our simulations (utilizing both centered and non-centered treatment indicator) as an alternative to considering a centralized treatment indicator above combined with independence tests. 

Due to the intuitive interpretation of the score residuals as an individual treatment effect (when using the centered parameterization), these can also be used to assess variable importance (\textit{i.e.} step 3b from Section \ref{sec:intro}). The advantage of extracting the score residuals first and then modeling the residuals directly is that the more complex problem of modeling the outcome including main effects and interactions has been reduced to just modeling the score residuals versus potential treatment effect modifiers. In addition the score residuals can be modeled with flexible modeling approaches, also allowing for higher order interactions among the covariates (compared to pure regression models that only consider linear additive interaction terms). In the residual based approaches, we propose to use conditional random forest \citep{hothorn2006unbiased} to model the score residuals versus all covariates, and obtain the variable importance from the model. 

To illustrate some of the concepts in this section (such as model dependence of estimand and the centered treatment indicator) we consider a simple example.

\subsubsection{Illustration}

%% NOTE: code is under illustration_score_resid.R in this project

We simulated data-sets with $n=200$ individuals, two covariates $x_{1,i} \sim N(0,1)$, $x_{2,i} \sim N(0,1)$, treatment indicator $z_i$ randomly drawn without replacement from among a set of 100 ones and 100 zeros and outcome $y_i\sim N(\mu_{m,i},1)$ with $\mu_{1,i}=1+3x_{1,i}+3z_i$ (no TEH) and $\mu_{2,i}=1+3x_{1,i}+3z_i+3z_ix_{1,i}$ (heterogeneous).

Based on this, six models are fitted
\begin{enumerate}
    \item[(M1)] Using the centered treatment indicator $\Tilde{z}_i=z_i-0.5$ and adjusting for the covariate $x_1$
    \item[(M2)] Using the centered treatment indicator $\Tilde{z}_i=z_i-0.5$ and adjusting for the wrong covariate $x_2$, omitting $x_1$ in the model
     \item[(M3)] Using the centered treatment indicator $\Tilde{z}_i=z_i-0.5$ and adjusting for both $x_1$ and $x_2$ in the model
    \item[(M4)] Using the non-centered treatment indicator $z_i$ and adjusting for the covariate $x_1$
    \item[(M5)] Using the non-centered treatment indicator $z_i$ and adjusting for the wrong covariate $x_2$, omitting $x_1$ in the model
    \item[(M6)] Using the non-centered treatment indicator $z_i$ and adjusting for both $x_1$ and $x_2$ in the model
    
\end{enumerate}

The models are fitted to two data-sets, the first data-set is produced using $\mu_{1,i}$, \textit{i.e.} no treatment effect heterogeneity. The second data-set is produced using $\mu_{2,i}$, that is when there is treatment effect heterogeneity.
\begin{figure}
    \centering
    \includegraphics[width=1\linewidth]{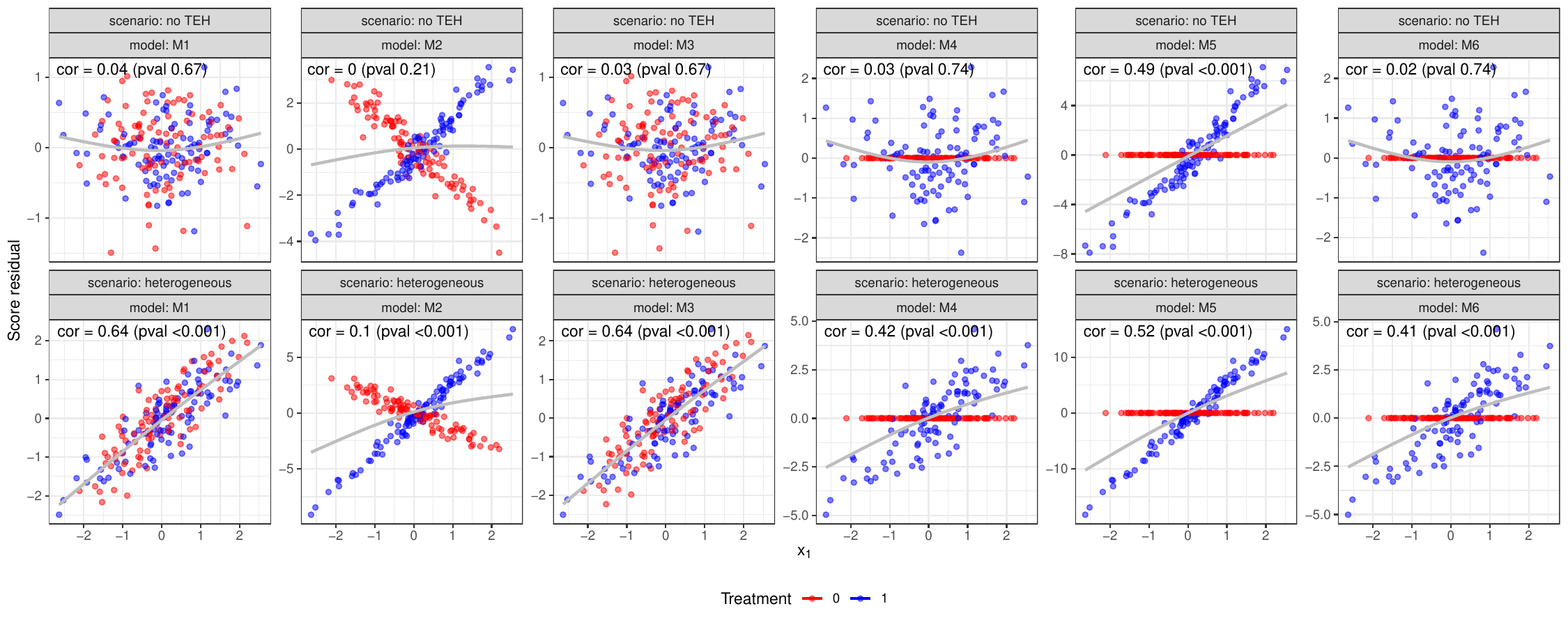}
    \caption{Plot of score residuals for models fitted to two data-sets versus the covariate $x_1$, corresponding to situations of no TEH and heterogeneous treatment effects. Model M1-M6 is in their corresponding columns. Kendall's correlation of $x_1$ and the score residual as well as the $p$-value for a global heterogeneity test are shown in the top left. The red points (treatment = 0) and blue points (treatment = 1) are the score residual for different treatment group, and the gray curve is the non-linear fitting of score residual versus $x_1$.}
    \label{fig:score_resid} 
\end{figure}
Figure \ref{fig:score_resid} shows the score residuals under the different models and situations, plotted against the variable $x_1$, which is prognostic and, when there is TEH, also predictive (is a treatment effect modifier). 

The performance of M3 and M6 is similar to M1 and M4, respectively (for both correlation and p-value), although M3 and M6 had an additional covariate $x_2$ (which is neither prognostic nor predictive). 

For model M1 and M4, when adjusting for the covariates from the data-generating model, the score residuals show no systematic heterogeneity in the setting of no TEH (Kendall's correlations of 0.04 and 0.03, with large $p$ values) and detect heterogeneity when it exists (correlation 0.64 and 0.42 and low $p$ values). 

Model M2 does not adjust $x_1$ as prognostic, but is still able to identify the case when there is no TEH and when there is TEH. Even though the results in both scenarios are worse, indicated by a lower p-value in the case of no TEH and lower correlation for the case of TEH compared to models M1 and M4.

Model M5 fails in the case of no TEH. It indicates high correlation in both the no TEH and TEH scenario. In the scenario of no TEH, the prognostic effect of $x_1$ induces a structure in the score-residuals for the treatment effect that indicate that $x_1$ is predictive.

There are two main learnings to take from this: (i) Even though there may be variables that are truly predictive (that is treatment effect modifiers that don't have a prognostic effect), in practice it is typically advisable to define the estimand and thus fit the model adjusting for all potential effect modifiers as prognostic (as suggested in Section \ref{sec:treatm-effect-meas}). (ii) It appears that usage of the centered treatment effect in the model leads to some robustness against mis-specification of the prognostic effects. This is likely due to the fact that the information about the outcome under both treatments is captured by the treatment effect coefficient in the centered parameterization and less independent of the intercept coefficient. Interestingly the overall fit of model $m$ is unaffected by how the treatment effect is parameterized, however the score residuals for the two parameterizations differ and the score residuals based on the centered treatment indicator contain more useful information for assessing TEH.

\section{Simulations}
\label{sec:simulations}

\subsection{Aims}

In this section, we will compare the performance of several versions of the residual based approach with alternative approaches mentioned in Section \ref{sec:global-teh-tests} on different types of outcomes, including continuous, binary, count and time-to-event outcomes.

% \noindent \textbf{Data generating mechanisms}\\
\subsection{Data Generating mechanisms}

For data generation we extended the scenarios from Sun et al.\cite{sun2024comparing}, available in the \textsf{R} package \textbf{benchtm}\cite{benchtm}. Originally the scenarios only considered continuous and binary outcome data. For the simulations here we extended the \textsf{R} package to include time-to-event and count outcomes. For self-sufficiency we will summarize the simulation scenarios, while for full details see the original publication.\cite{sun2024comparing} We consider the setting of a two-arm randomized study where the response is generated from
\begin{equation}
\label{form:data-generation}
\eta_i = f_{prog}(\mathbf{x}_i) + z_i(\gamma_0 + \gamma_1 f_{pred}(\mathbf{x}_i)),
\end{equation}
where $f_{prog}(.)$ is a function that captures the prognostic effects and $f_{pred}(.)$ the predictive effects modifying the treatment effect, and $z_i \in \{0,1\}$ is the treatment indicator.

The outcome data $y_i$ are generated in the following way.
\begin{itemize}
    \item For continuous outcomes: $y_i \sim N(\eta_i,\sigma^2)$ 
    \item For a binary outcome: $y_i \sim B(g(\eta_i))$, for the Bernoulli distribution $B(.)$ and $g(x)=1/(1+\exp(-x))$.
    \item For count data: $y_i \sim NB(\exp(\eta_i),\theta)$, where $NB(\mu,\theta)$ is the negative binomial distribution with mean $\mu$ and dispersion parameter $\theta$ (so that the variance is $\mu(1+\mu/\theta)$).  
    \item For time-to-event outcomes we observe $(y_i,c_i)$, where $y_i$ is the time and $c_i$ indicates whether the event is observed ($c_i=1$) or censored ($c_i=0$). $y_i=min(\tilde{y}_i,\tilde{t}_i)$, where $\tilde{y}_i \sim Exp(\lambda_0\exp(\eta_i))$ is the exponential distribution and $\tilde{t}_i\sim \mathcal{C}$. For details on $\lambda_0$ and $\mathcal{C}$ see Appendix \ref{sec:app_tte}.
\end{itemize}

A sample size of $n = 500$ is utilized with $k=30$ baseline covariates. The treatment indicator is generated by permuting a vector with 250 zeros and ones. The covariates are generated as standardized synthetic covariates from a real clinical trial, so mimicking a typical covariance structure across clinical covariates. For $f_{prog}(\mathbf{x})$ and $f_{pred}(\mathbf{x})$ four scenarios are considered in Table \ref{tab:sim_models}. As the magnitude of prognostic effects on the outcome (and relative size of prognostic versus predictive effects) can impact the performance, we selected the scaling factor $s$ in Table \ref{tab:sim_models} so that realistic $R^2$ or $AUC$ values are achieved on the control arm. For each outcome type we selected these values based on different actual clinical trials: For continuous data an $R^2=0.32$ was chosen, for binary data an $AUC=0.66$ was chosen\cite{sun2024comparing}. For the new scenarios of negative binomial $R^2=0.41$ and for time-to-event data $R^2=0.32$ (Cox and Snell $R^2$ \cite{allison2013s}) was used, based on the utilized studies. For these two cases the $R^2$ was defined as proportion of randomness/deviance explained\cite{o2005explained}.

The parameter $\gamma_1$ is the major factor that determines how challenging it is to detect treatment effect heterogeneity. Five scenarios are considered: one is $\gamma_1=0$ where there is no treatment effect heterogeneity. Then the value $\gamma^*_1$ was calculated that leads to a power of 0.8 for the interaction test for $H_0: \gamma_1=0$ vs $H_1: \gamma_1 \neq 0$ under the true data-generating model from \eqref{form:data-generation}, for a Type I error of 0.1. In the simulations then these four additional scenarios are considered $0.5\gamma^*_1$, $\gamma^*_1$, $1.5\gamma^*_1$ and $2\gamma^*_1$. The parameter $\gamma_0$ is chosen so that a power of 0.5 is achieved for detecting positive overall treatment effect (with one sided $\alpha = 0.025$) for a given $\gamma_1$. This power is a realistic setting in a Phase III study, where the overall treatment effect is slightly smaller than that planned for at the design stage. 

\begin{table}[h]
\centering
\caption{Simulation models. Here $a\lor b$ means ``a or b'', $a\land b$ represents ``a and b'', $I(.)$ is the indicator function, $\Phi$(.) is the cumulative distribution function (cdf) of standard normal distribution, and $s$ is a scaling factor that is chosen depending on the simulation scenario to achieve a specific $R^2$ or $AUC$ on the control group. 'Y', 'N' are two different levels from the categorical variable $x_1$, $x_4$ and $x_8$. Please check \cite{benchtm} for more details on the data generation.} 
\label{tab:sim_models}
 \begin{tabular}{l l} 
 \hline
Scenario & $\eta_i$ \\ \hline
1 & $s\times (0.5I(x_{1,i}=\text{'Y'})+x_{11,i}) + z_i(\gamma_0 + \gamma_1\Phi(20(x_{11,i}-0.5)))$ \\
2 & $s\times (x_{14,i}-I(x_{8,i}=\text{'N'})) + z_i(\gamma_0 + \gamma_1x_{14,i})$ \\ 
3 & $s\times (I(x_{1,i}=\text{'N'})-0.5x_{17,i}) + z_i(\gamma_0 + \gamma_1I((x_{14,i}>0.25)\land (x_{1,i}=\text{'N'})))$ \\ 
4 & $s\times (x_{11,i}-x_{14,i}) + z_i(\gamma_0 + \gamma_1I((x_{14,i}>0.3)\lor(x_{4,i}=\text{'Y'})))$ \\ \hline
\end{tabular}
\end{table}

% \noindent \textbf{Model based method options}
\subsection{Compared methods}
\label{sec:methods}

\subsubsection{Residual-based methods}
\label{sec:resid_based}

The residual-based method described in Section \ref{sec:methodology} is implemented in different variations. For the independence test between score residuals and other covariates, the permutation independence test as described in Section \ref{sec:using-score-residual} is used, as implemented in the function \texttt{independent\_test} from the \textbf{coin} package \cite{coin2008}. 
% We will compare maximum and quadratic test statistic\cite{hothorn2006lego}. If the predictive effect is driven by a single important variable, one would expect the maximum test to perform better, whereas if multiple variables jointly influence the treatment effect, one would expect the quadratic test statistic to perform better. 
The considerations in Section \ref{sec:using-score-residual} suggest that a centered treatment parameter may perform better than a non-centered treatment effect indicator. Hence we will use centered treatment indicator in our following comparison.
% Hence we will compare both options in the first simulation study. 
In Section \ref{sec:treatm-effect-meas} it was discussed that we target the estimand where all relevant potential treatment effect modifiers are adjusted for as prognostic effect. This however does not specify how the parameters of the prognostic effects should be estimated. After comparison of different prognostic variable selection options (see Section \ref{sec:result_residual} for details), we use LASSO and ridge regression for the prognostic effect estimation. In the following section, we compare alternative methods with two residual score approaches—one using LASSO and the other ridge regression for estimating prognostic effects—both applying a maximum test with a centered treatment indicator. These are referred to as \textit{Residual LASSO} and \textit{Residual Risk}.

\subsubsection{Alternative approaches}
\label{sec:altern-appr}

The residual based approaches will be compared to four main alternative methods. All alternative methods will be fitted in a two-stage manner, where first a selection of the prognostic effects will be done based on the LASSO method as described in the previous section. Then the corresponding method will be performed utilizing the identified covariates as prognostic effects.

A likelihood ratio test will be implemented as described in Section \ref{sec:global-teh-tests}, we will utilize the likelihood ratio test statistic and the exact critical value for normally distributed data. For the other outcome types the asymptotic $\chi^2$ critical value will be used (this method will be denoted as \textit{Likelihood Asymptotic}). In addition a parametric bootstrap test will be performed as described in Section \ref{sec:global-teh-tests} (this method will be denoted as \textit{Likelihood Bootstrap}). See Appendix \ref{sec:boot_tte} for how the bootstrapping is performed for time-to-event data. For assessing variable importance only \textit{Likelihood Asymptotic} is included. Variable importance will be assessed for each variable by performing a likelihood ratio test comparing the models with and without including the corresponding variable as interaction (while always still including all prognostic effects and all other interactions), the variables are then ordered according to the resulting $p$-value.

In addition Goeman's global test (denoted as \textit{Goeman's Test}) will be implemented using the \textbf{globaltest} \textsf{R} package, with the same selected variables as prognostic effects, and all covariates separately interacting with treatment indicator. For continuous data the \texttt{"linear"}, for binary data the \texttt{"logistic"} and for survival data the \texttt{"cox"} option will be used. For count outcome data, we generate data from a negative binomial distribution, which is not available in the \textbf{globaltest} package, here we will use the \texttt{"poisson"} option. To assess predictive variable importance we will utilize the variable importance measures as available in the package via the \texttt{covariates} function.

Finally we will use the model-based-partitioning (MOB) method \cite{zeil:hoth:horn:2008} for the global test. The LASSO selected variables are adjusted for as prognostic effects. We use the \texttt{mob} function from the \textbf{partykit} package for this purpose and select the \texttt{parm} option in a way so that only the residual scores with respect to the treatment effect parameter are used in the test. For performing a global test it is enough to assess the test statistics at the root node of the tree-building process. For each of the 30 covariates a fluctuation test is performed and the resulting $p$-values are adjusted based on the Bonferroni method. In Sun et al. (2024)\cite{sun2024comparing} this method was among the best methods for the global heterogeneity test for normally distributed and binary data. The MOB approach will also be utilized both with centered and non-centered treatment indicator. We will abbreviate the corresponding methods as \textit{MOB centered} and \textit{MOB non-centered}. We don't compare the MOB methods in terms of its performance for predictive variable importance as there is no obvious way on how to obtain multivariate/adjusted variable importance. 
Note that the two MOB approaches are essentially identical with the residual based approaches, as they are also based on testing for dependency of the score residual with covariates. There are two differences: (i) in MOB the residual scores are orthogonalized before the test is applied to the column corresponding to the treatment effect parameter and (ii) in MOB the utilized test is not an independence test, but a univariate fluctuation test for each covariate (requiring usage of the Bonferroni correction)\cite{zeil:hoth:horn:2008}.

As a comparison, the Oracle method is also added where the true prognostic and predictive effect are incorporated into the model with true form this will be abbreviated as \textit{Oracle} in what follows. Here the interaction test can be done by just testing for $\gamma_1=0$.

% \noindent \textbf{Performance Measures}\\
\subsection{Performance Measures}
\label{sec:performance}

The following measures will be utilized to compare the performance of the different methods.

\begin{description}
    \item[Measure 1:] The distribution of the $p$-value resulting from the corresponding global TEH test. In the scenario of homogeneity ($\gamma_1 = 0$) one would expect, that the distribution is a uniform distribution on the interval $[0,1]$, so that the empirical distribution function (ECDF) of the p-values should be close to the straight diagonal line increasing from 0 to 1. Under homogeneity a deviation towards systematically larger or smaller values compared to a uniform distribution, would be undesired as it would result in conservative or anti-conservative $p$-values for the scenario. Under the situation of heterogeneity the $p$-value distribution should be shifted towards 0 for better methods. The performance will be assessed visually with ECDFs of the $p$-value distribution and under the alternative, when comparing to other methods, by visualizing the median surprise value ($-\log_2(\mathrm{p-value})$) .
    \item[Measure 2:] To assess variable importance ranking in the setting of homogeneous treatment effects, one would expect that each of the 30 variables is selected equally frequently (that is with probability 1/30) at the top of the variable importance ranking. In the case of heterogeneous treatment effects we consider whether the variable identified as the top predictive variable is truly one of the predictive variables for this scenario. 
\end{description}

\subsection{Results}
\label{sec:results}
Simulation under all scenarios were conducted using 500 repetitions. For each simulation all methods were applied to the same simulated data to enhance the precision when comparing across methods. 
% \subsubsection{Comparison versus alternative methods}

\begin{figure}[htbp]
    \centering
    \includegraphics[width=0.7\linewidth]{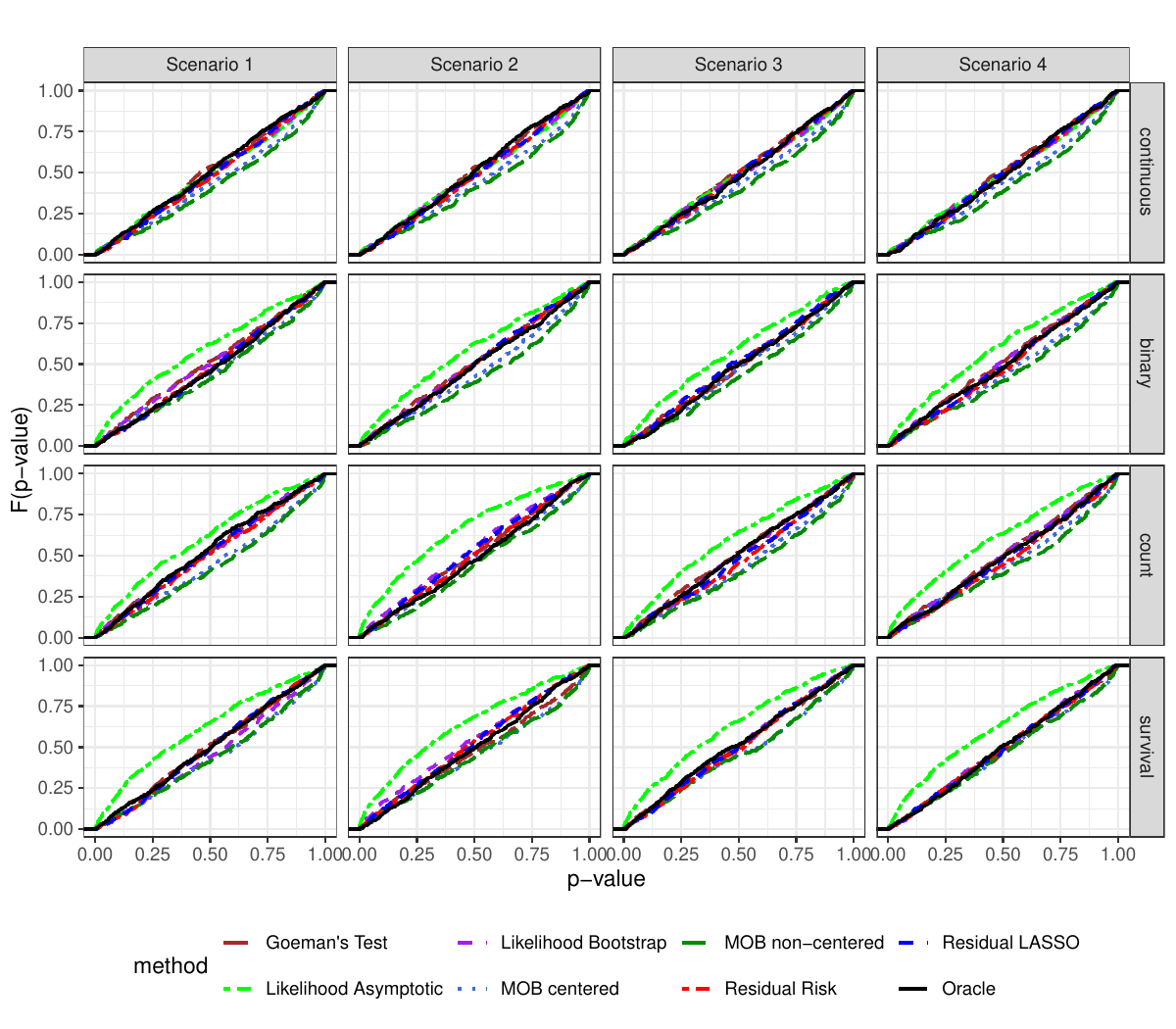}
    \caption{ECDF for $p$-values (F(p-value)) for each method under null hypothesis, i.e. no heterogeneity. A good method should have $p$-value uniformly distributed along 0, 1, with an ECDF line follows straight diagonal line.}
    \label{fig:pvalue_null} 
\end{figure}

When there is no TEH, the $p$-value distribution of each method is displayed in Figure \ref{fig:pvalue_null}. It seems only \textit{MOB centered} and \textit{MOB non-centered} and \textit{Likelihood Asymptotic} deviate from the uniform distribution. For the non-continuous settings the \textit{Likelihood Asymptotic} method has ECDF above diagonal line, which indicates the existence of $p$-values that are too small for a uniform distribution. This may happen because the chi-squared approximation used for the likelihood ratio test statistic in the non-continuous setting fails. This observation is in line with Sur et al's exploration on binary case \cite{sur2019likelihood}. The method is hence anti-conservative and should not be used as the $p$-values are not calibrated and not easily interpretable. In addition, MOB appears to have a slightly conservative $p$-value distribution (with ECDF curve below diagonal line), which is likely related to the Bonferroni adjustment that is used. 

\begin{figure}[htbp]
    \centering
    \includegraphics[width=0.7\linewidth]{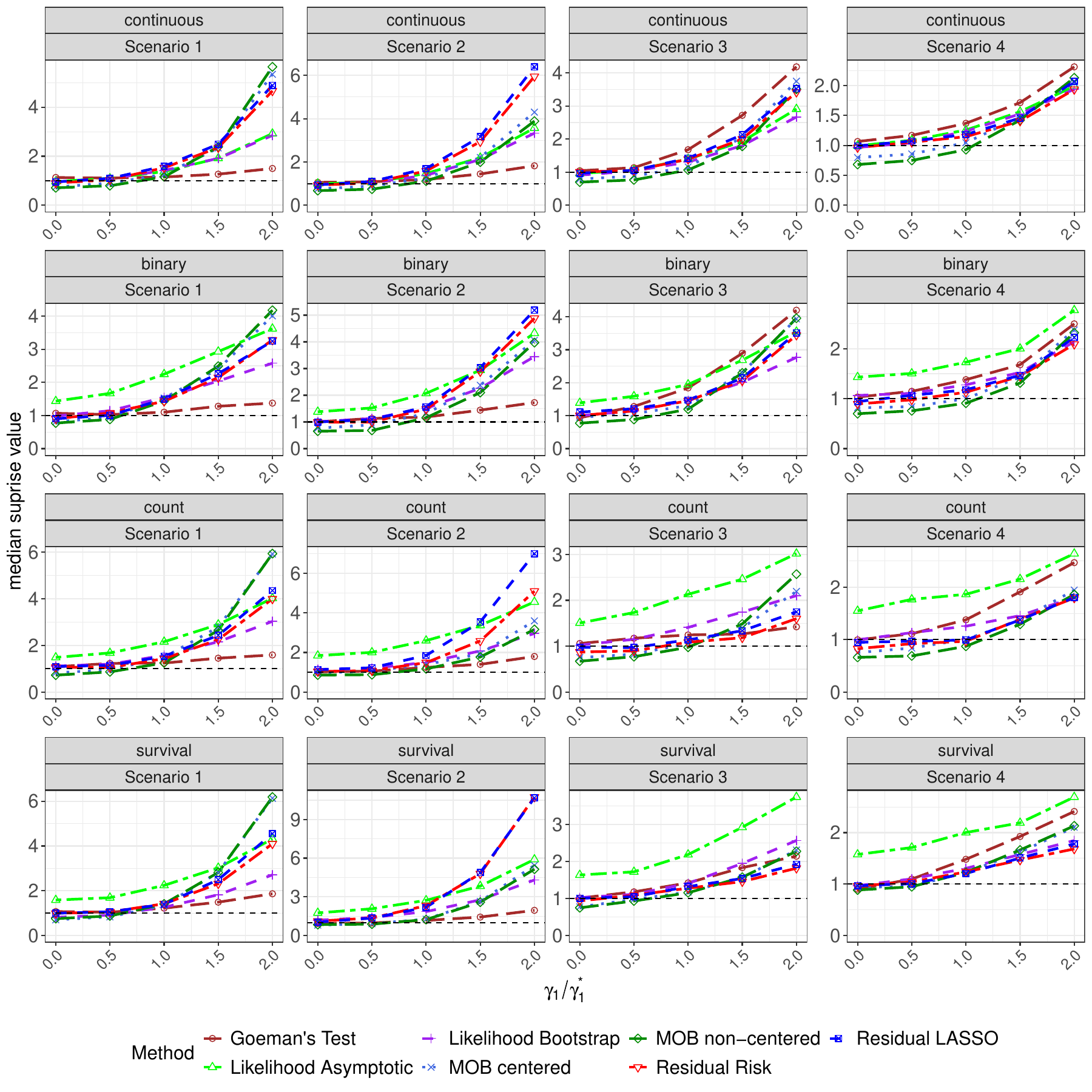}
    \caption{Median surprise value (-log2($p$-value)) under various treatment effect heterogeneity, a large median surprise value is corresponding to a large power to detect treatment effect heterogeneity. The black dashed line is representing surprise value of 1, which is corresponding to $p$-value of 0.5.}
    \label{fig:pvalue_alt} 
\end{figure}

When there is TEH, the median surprise value, calculated as -log2($p$-value) is displayed for each method under different scenarios in Figure \ref{fig:pvalue_alt}. When there is no TEH ($\gamma_1/\gamma^*_1 = 0$), one would expect the median surprise value to be -log2(0.5)=1 (black dashed line). As the amount of heterogeneity increases the median surprise value increases for all methods. \textit{Likelihood Asymptotic} in general has large surprise value in all scenarios, though its results are difficult to interpret as it median surprise values $>1$ under the scenario with no TEH (except for continuous endpoints). The performance of \textit{Goeman's Test} compared to the residual methods is case dependent, for Scenario 1 and 2, \textit{Goeman's Test} performs worse compared to residual based method, but slightly better on Scenario 3 and 4. In general, MOB performs slightly worse compared to residual based method, especially on the case where predictive variable is continuous (X14), and treatment effect heterogeneity is small ($\gamma_1/\gamma^*_1 <= 1$). The residual based method has decent performance when there is TEH, and the performance looks quite similar no matter whether LASSO variable selection or risk is used as prognostic effect.

\begin{figure}[htbp]
    \centering
    \includegraphics[width=0.7\linewidth]{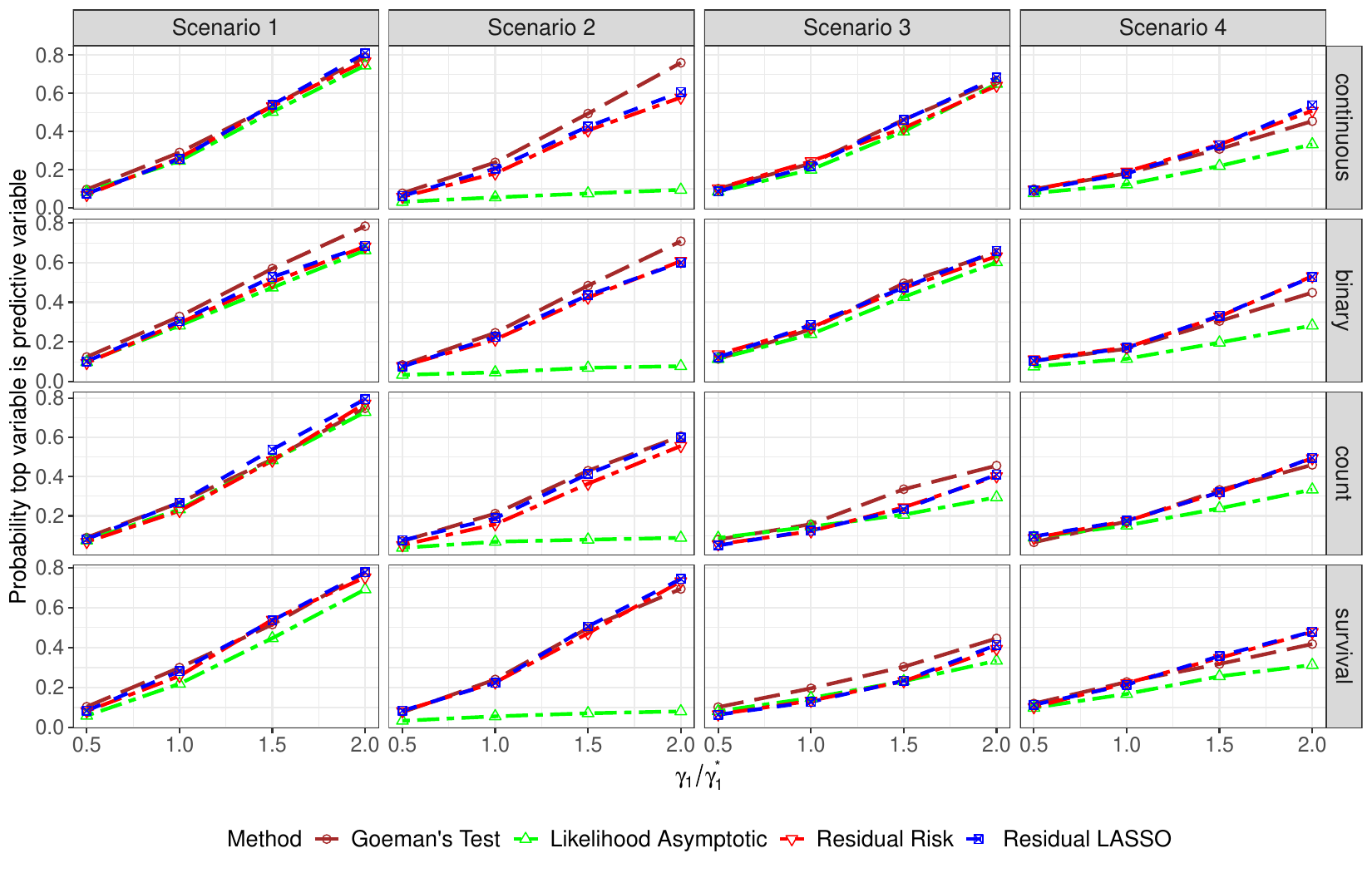}
    \caption{Probability of top selected variable from variable importance being predictive}
    \label{fig:top1}
\end{figure}

To assess the quality of the variable importance ranking in the simulation study, Figure \ref{fig:predprob} in the Appendix displays how often each variable is selected. Under the scenario of homogeneous treatment effect when there is no predictive variable, each variable should be at the top of the ranking with an equal probability of 1/30 (as there are 30 variables in total). It can be seen that all methods perform well in this metric, except for \textit{Goeman's Test}, that has frequency selected a categorical variable $X_3$ in all scenarios. The potential reason is $X_3$ is a categorical variable with 5 levels, which is coded into 4 dummy variables. In addition 2 of 5 variables have observations $< 10\%$ of total sample size, and the variable importance calculated from \textit{Goeman's Test} for categorical variable is based on each level, and a level with small size might be more likely to have enhanced/decreased treatment effect compared to complement. 

With TEH existence we evaluate how often the top ranked variable is actually predictive on Figure \ref{fig:top1}. It can be seen that depending on the scenario either the \textit{Goeman's Test} or the residual based approach work best. The standard approach of assessing variable importance based on regression models and likelihood ratio tests is consistently outperformed and not recommended.

To summarize the results, the residual based methods (with centered treatment indicator) both with risk and LASSO-selected prognostic effects work well with slight advantages for the LASSO based approach. The residual based approach compares similar to the MOB based tests. Interestingly for the MOB based tests there is no major difference between centering or not centering the treatment indicator (which was also observed in Dandl et al. (2024)\cite{dandl2024makes} for data from randomized trials). We think this can at least partially be explained by the fact that the score residuals are orthogonalized before performing the test. \textit{Goeman's Test} also leads to very competitive performance in many simulation scenarios, and may have benefits in particular in higher-dimensional situations (which we did not consider in our simulations). Note, however, that residual-based approach (via use of the conditional random forest for assessing variable importance) may be able to deal with interactions better than the \textit{Goeman's test}, which includes the covariates just as linear additive parameters on the treatment effect. The standard likelihood ratio test is only recommended when implemented in a bootstrap testing approach (for non-normal data). Variable importance assessment from regression models cannot be recommended as they are consistently outperformed.

\section{Example}
\label{sec:example}

To illustrate the methods in this article, we will use synthetic simulated data that mimic data from a cardiovascular outcome study, which evaluated an investigational drug versus placebo with time to MACE (major adverse cardiovascular events) as the primary endpoint. In the Appendix we include an extra example considering the same dataset but  with continuous endpoints and multiple dosages treatments. The data contains 3344 patients on the placebo arm and 2263 on the treatment arm. 

\begin{figure}[htbp]
    \centering
    \includegraphics[width=0.4\linewidth]{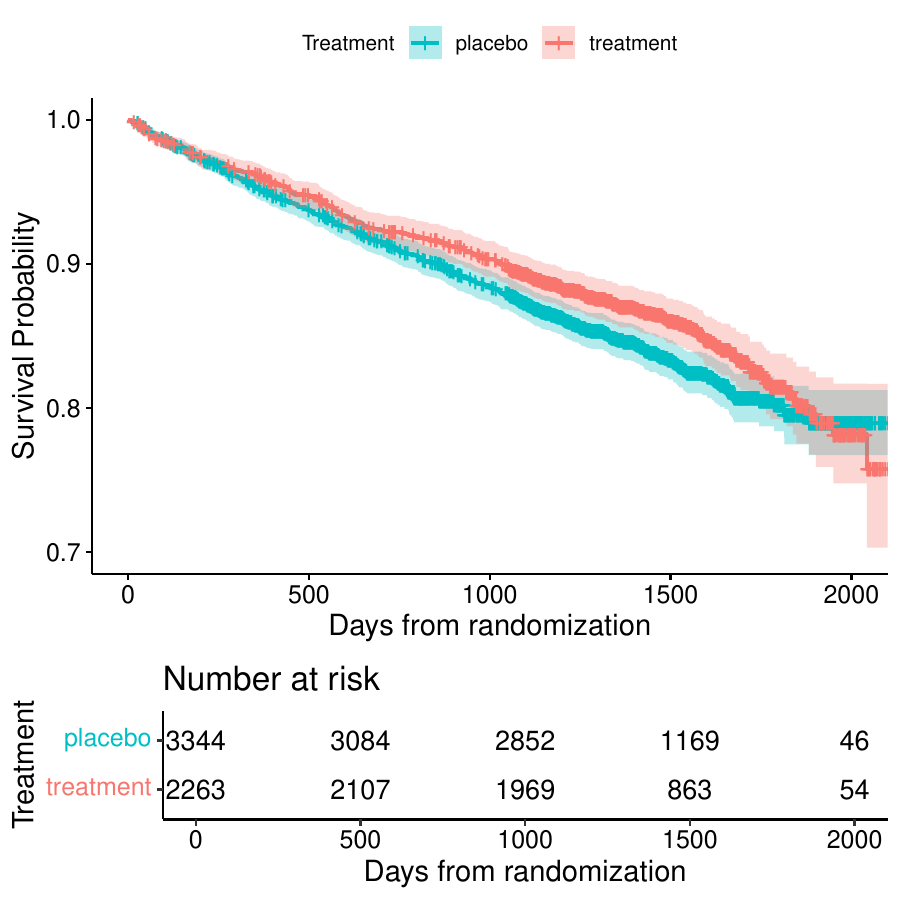}
    \caption{Kaplan-Meier plot for for event free probability over time for different treatment arms. Red curve is for treatment arm and green curve is for control arm.}
    \label{fig:km} 
\end{figure}

Figure \ref{fig:km} shows the Kaplan-Meier curve for event free probability curve over time for different treatment arms. Overall, there is difference between the event free probabilities for treatment and placebo arm, and treatment arm tends to have higher probability of free event compared to placebo arm in early years (p-value = 0.036 from Cox model).

In this example, the interest is to explore the treatment effect heterogeneity for log hazard ratio, based on the proportional hazard model 
\begin{equation}
  \label{eq:hr_mod}
    \log(h_i(t)) = \log(h_0(t)) +\mathbf{x}_i'\boldsymbol{\beta}_\mathbf{x}+\delta (z_i-z_m)\;\mathrm{with}\;i=1,\ldots,N,
\end{equation}

where $z_i = 1$ if a subject receives treatment and $z_i = 0$ if receives placebo and $z_m = \frac{1}{N}\sum_{i=1}^N {z}_i$. The treatment effect (treatment versus placebo) in this model is hence given by $\delta$. To explore treatment effect heterogeneity 19 potential effect modifiers will be utilized, described in detail in Table \ref{tbl:variables} in the Appendix. These variables were selected based on the subgroup variables pre-defined for subgroup analyses for the primary endpoint. The vector $\mathbf{x}$ is hence an 19-dimensional vector for each patient, containing the potential effect modifiers. 

We fitted the model \eqref{eq:hr_mod} in two variations, V1: including all 19 covariates as well as treatment in the Cox model. V2: first fitting a penalized Cox regression without treatment indicator, using a ridge penalty and choosing the penalty parameter according to 10-fold cross-validation. Based on the penalized model then a model is fitted adjusting for treatment (as in equation \eqref{eq:hr_mod}) and the resulting risk score as prognostic effect. For both versions, the score residual for treatment indicator is extracted. In addition we will utilize the MOB methodology with a non-centered treatment indicator adjusting in the same way as described in V1 and V2, and extract the $p$-value for the fluctuation test in the root node.

When using permutation independence test as a global test on treatment versus all covariates with maximum test statistics, the test results in $p$-values of 0.88 (V1) and 0.89 (V2). On the surprise scale (-log2(p)) these correspond to 0.19, 0.16, which according to the scale in \cite{sechidis2024watch} corresponds to weak evidence against the null-hypothesis of treatment effect homogeneity. The result is quite consistent compared to that of MOB which gives a $p$-values of 0.87 (surprise value 0.20) using all covariates (V1) and 0.70 (surprise value 0.51) using risk as covariate (V2). 

Variable importance measures have been calculated by fitting a conditional random forest \cite{hothorn2006unbiased} to the score residuals (based on the model fit \eqref{eq:hr_mod}) and calculating the permutation importance of the underlying variables (see Figure \ref{fig:vi}). As can be seen the variable importance assessment is quite variable, and the difference between different variables are not large, which is consistent with the finding of little evidence against homogeneity. The variable importance plot from both versions are quite consistent. We further investigated Kaplan-Meier plot and log hazard ratio in Appendix based on top variable selected ethinic and baseline mean sitting diatolic BP (mmHg)(VSSTDBMB), and did not find much treatment effect differences on different levels of the variables.  

\begin{figure}[htbp]
    \centering
    \subfloat[Variable importance for V1]{%
        \centering
        \includegraphics[width=0.35\textwidth]{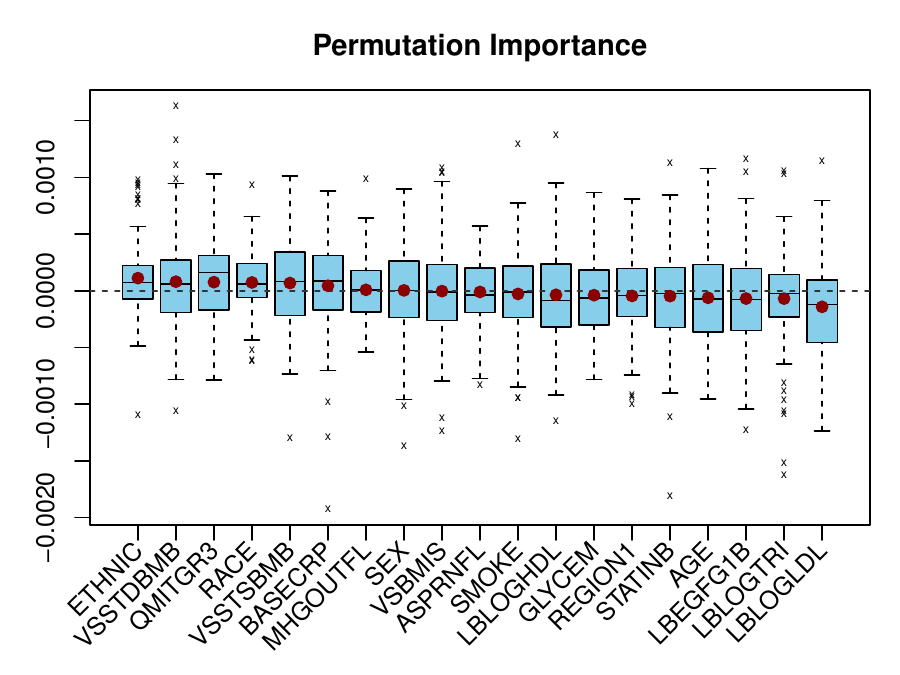}
    }
    \subfloat[Variable importance for V2]{%
        \includegraphics[width=0.35\textwidth]{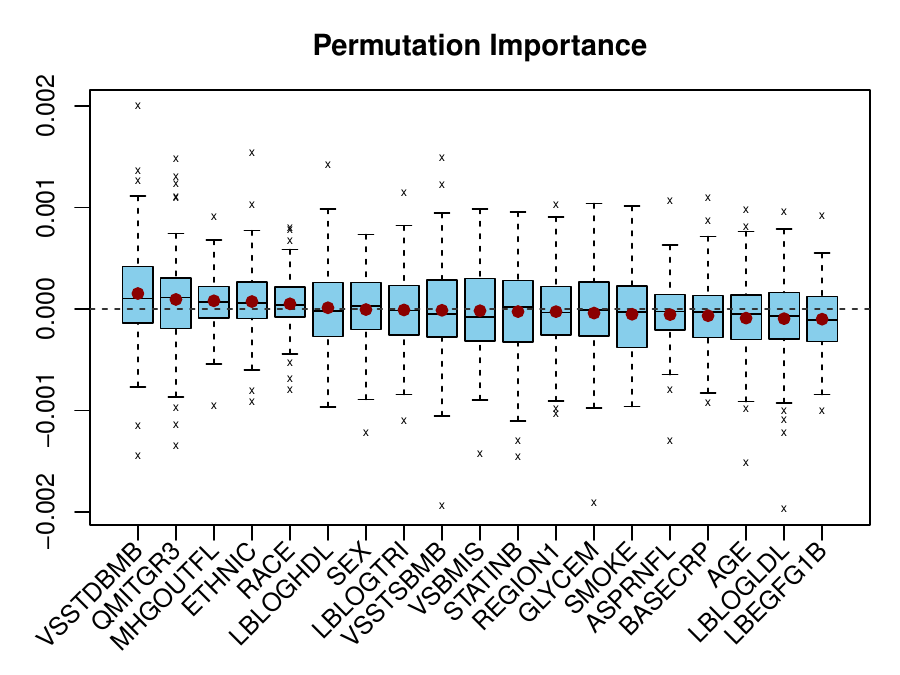}
    }
    \caption{Variable importance for treatment effect modifiers for two versions of models. For each boxplot, each point is the permutation importance for each tree based on conditional random forest with 100 tress.}
    \label{fig:vi}
\end{figure}

\section{Conclusions}
\label{sec:concl}

In this work we illustrate how to assess treatment effect heterogeneity in situations where the treatment effect is implicitly defined by usage of regression models. This is an important application as in current clinical research practice regression model-based effect measures are still commonly utilized, in particular for time-to-event and binary outcomes. 

To this end, we review in detail (i) the notion of treatment effect heterogeneity and that it depend on the utilized effect measure (and thus selection of the effect measure is important) and (ii) how treatment effect measures are implicitly defined, when a regression model is utilized (in Section \ref{sec:treatm-effect-meas}). Based on this then different methods to implement the WATCH workflow are described, ranging from standard global interaction tests in regression models (using asymptotic and bootstrap critical values or methods utilizing ideas from random effects modeling) to different methods that utilize the score residual with respect to the treatment effect parameter, building on earlier work\cite{hothorn2006unbiased, zeil:hoth:horn:2008, seib:zeil:hoth:2016, thomas2018subgroup}. 

If appropriately utilized, the score residual with respect to the treatment effect parameter can be interpreted as an individual deviation from the overall treatment effect. In that sense similar methods for global heterogeneity tests and assessment of variable importance can be used as in the approach presented in Sechidis et al. (2025) \cite{sechidis2025using}, which uses doubly robust pseudo-outcomes for the treatment effect for each patient in the trial. More specifically an independence test and usage of the permutation variable importance underlying conditional random forest seem to work well across both situations. Even though other methods may be used as well.

In our simulations we then investigated different versions of the residual-based approach. Based on the results a centered treatment effect indicator should be used, furthermore the maximum test statistic should be used if it is expected that heterogeneity is driven by a low number of variables, while the quadratic test statistic is more suited when several variables impact the results.

We also compared these approaches with further methods illustrating that standard global interaction tests based on regression models and using the asymptotic chi-squared distribution cannot be recommended, for either the global heterogeneity test or the variable importance measure. In terms of global tests the methods based on the residual scores (including MOB) performed well, but also the \textit{Goeman's Test} approach and no method uniformly outperformed the other. For MOB it did not matter whether or not the treatment effect is centered, which is likely due to the fact that the residual scores are orthogonalized before performing the test.

We also investigated the score residual approach in a synthetic trial example with time-to-event endpoint, where all evaluated methods show consistent findings.

We propose to define the target estimand always by adjusting for all possible effect modifiers, but suggest to use penalized estimation methods for prognostic effects, such as based on a variable selection (i.e. allowing some parameters to be exactly 0) or shrinkage estimation, in particular if a larger number of covariates is utilized. For that purpose, across the scenarios the LASSO slightly outperformed the risk based approach (based on ridge regression). This may partially due to our simulation settings, in which only two variables were prognostic in each case. There is one practical consideration that speaks for utilizing the risk approach based on ridge: In practice the tuning parameter in the shrinkage penalty is chosen by cross-validation, which involves a random element. Here we saw in practical applications that the overall results (e.g., heterogeneity $p$-value) based on LASSO are more affected by this randomness (and thus less stable) than the risk approach based on ridge. 

\section*{Acknowledgments}
The authors would like to thank EFSPI Treatment Effect Heterogeneity Special Interest Group (SIG) for their many discussions and feedback.

\section*{Data availability statement}
Due to confidentiality and privacy concerns, the patient level data from the Example section cannot be shared publicly. The code used for simulation and analysis will be available at: https://github.com/Novartis/WATCH/.

\appendix

\section{Appendix}
\label{sec:appex}

\subsection{Details on data generation for time-to-event}
\label{sec:app_tte}
\begin{figure}[htbp]
    \centering
    \includegraphics[width=0.6\textwidth]{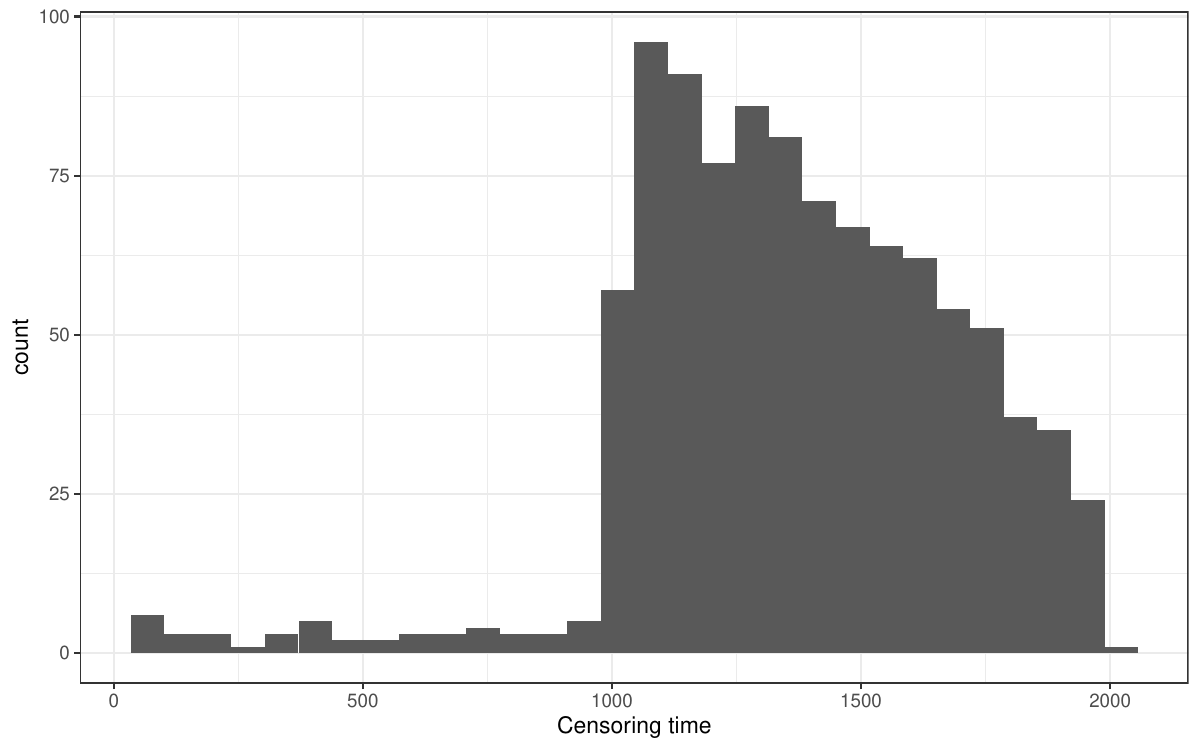}
    \caption{Distribution of censoring (days) for n = 1000}
    \label{fig:cens}
\end{figure}

For time-to-event endpoints, the censoring time $\mathcal{C}$ is generated as a mixture distribution of uniform distribution (before cutoff 1000) and beta distribution (after cutoff 1000 and before end of study time 2000). This setting is based on clinical trial setting with 3 year (around 1000 days) recruiting time and 3 year follow up time, during the follow up, patient randomly drop out of the study with uniformly distributed drop out time, at the end of the study (last patient reach 3 year follow up), everyone without an event will be censored with different follow up time (as they have different starting time). The function for generating censoring time is provided, as well as the distribution of censoring time (Figure \ref{fig:cens}) with n = 1000. $\lambda_0$ is chosen such that when there is no TEH, $P(\tilde{y} < 2000) < 0.5$.
\begin{lstlisting}
cens_time <- function(n) {
  p <- sample(0:1, n, replace = TRUE, prob = c(0.05, 0.95))
  r1 <- runif(n, 0, 1000)
  r2 <- 1000 + (2000 - 1000) * rbeta(n, 1, 1.5)
  (1 - p) * r1 + p * r2
}
\end{lstlisting}

\newpage

\subsection{Details on bootstrap approach for time-to-event data}
\label{sec:boot_tte}
In this section, the algorithm for bootstrapping for time-to-event endpoint is listed below.

\begin{enumerate}
    \item Cox model to model main effect $Surv(Y, event) \sim z + x_1 + x_2 + \ldots + x_k$, and main effect with interaction effects $Surv(Y, event) \sim z + x_1 + x_2 + \ldots + x_k + z*x_1 + z*x_2 + \ldots+z*x_k$ and obtain their partial likelihood difference $D$
    \item Model time to censor Mc using Cox model: $Surv(Y, event) \sim z + x_1 + x_2 + \ldots + x_k$, obtain probability of survival for each subject
    \item Model time to death Md using Cox model: $Surv(Y, event) \sim z + x_1 + x_2 + \ldots + x_k$
    \item for each bootstrap sample $j$, $j = 1, \ldots, 1000$:
    \begin{enumerate}
    \item generate censoring time for each subject $\tilde{t}_1, \tilde{t}_2, \dots, \tilde{t}_n$ from Mc
    \item generate survival time $\tilde{y}_i$ for each subject from Md
    \item observed death/censor time $y_i = min(\tilde{y}_i, \tilde{t}_i)$
    \item get the difference $D_j$ of the  partial likelihoods from a Cox model with main effect only versus a Cox model with main effects and interaction effects for each bootstrap sample 
    \end{enumerate}
    \item Calculate p-value $= P(D_j > D)$
\end{enumerate}

\newpage

\subsection{Additional figures}
\label{sec:add_fig}
\subsubsection{Variable selection probabilities when there is no TEH}

\begin{figure}[htbp]
    \centering
    \includegraphics[width=0.75\linewidth]{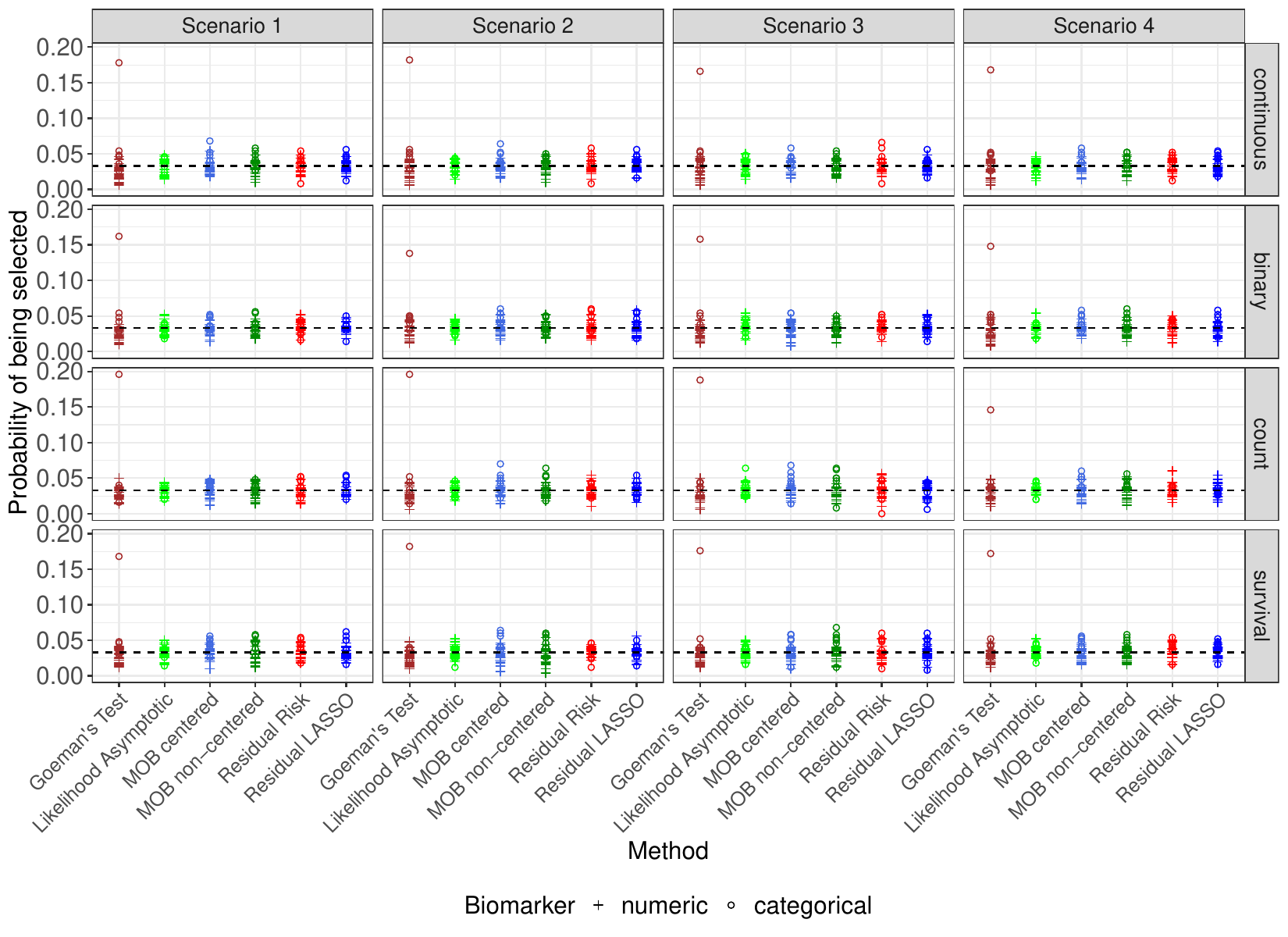}
    \caption{Probability each variable is being selected as top variable when there is no heterogeneity. A good method should have each variable being selected as predictive variables with equal probability, which is 1/30 (the dashed horizontal line).}
    \label{fig:predprob}
\end{figure}

\newpage

\subsection{Details on different options of model based methods}
\label{sec:result_residual}
In Section \ref{sec:resid_based}, we briefly introduced the different options we adopt for the independent test for model based methods. Here we demonstrate in more details with simulation to explain why these options are used. The residual-based method described in Section \ref{sec:methodology} is implemented in different variations. 
For the independence test between score residuals and other covariates, the permutation independence test as described in Section \ref{sec:using-score-residual} is used, as implemented in the function \texttt{independent\_test} from \textbf{coin} package \cite{coin2008}. 
The considerations in Section \ref{sec:using-score-residual} suggest that a centered treatment parameter may perform better than a non-centered treatment effect indicator. Hence we will compare both options in the first simulation study.
In Section \ref{sec:treatm-effect-meas} it was discussed that we target the estimand where all relevant potential treatment effect modifiers are adjusted for as prognostic effect. This however does not specify how the parameters of the prognostic effects should be estimated. Including all covariates as prognostic effects and using maximum (partial-)likelihood may overfit the observed data. Hence we also include variable selection (which may be interpreted as setting some of the estimated model parameters to 0) or penalized estimation methods, which shrink the parameter estimates towards 0. In total we will consider 4 approaches: (i) oracle: This knows the true prognostic and predictive effects and just estimates a parameter scaling the true prognostic and predictive effects. This method serves as benchmark for all other methods. (ii) all: An approach that just uses maximum likelihood (partial maximum likelihood for time-to-event data) and estimates the model adjusting for the treatment indicator (centered and non-centered) and all 30 covariates as prognostic effects. (iii) LASSO: A two-stage approach, where first a LASSO model is fitted to each treatment arm separately. The LASSO tuning parameter is selected by minimizing cross-validation loss. Then the combined selected covariates are used as prognostic effect in a model that also includes a treatment indicator. (iv) risk: In this approach in the first stage a ridge regression is fitted to all data, but without a treatment indicator variable (so modelling the overall average across treatment and control group). The tuning parameter is selected by minimizing cross-validation loss. The derived risk score (on linear scale) is then used as the only prognostic effect in the second stage model that also includes a treatment indicator.

\begin{figure}[htbp]
    \centering
    \includegraphics[width=0.7\linewidth]{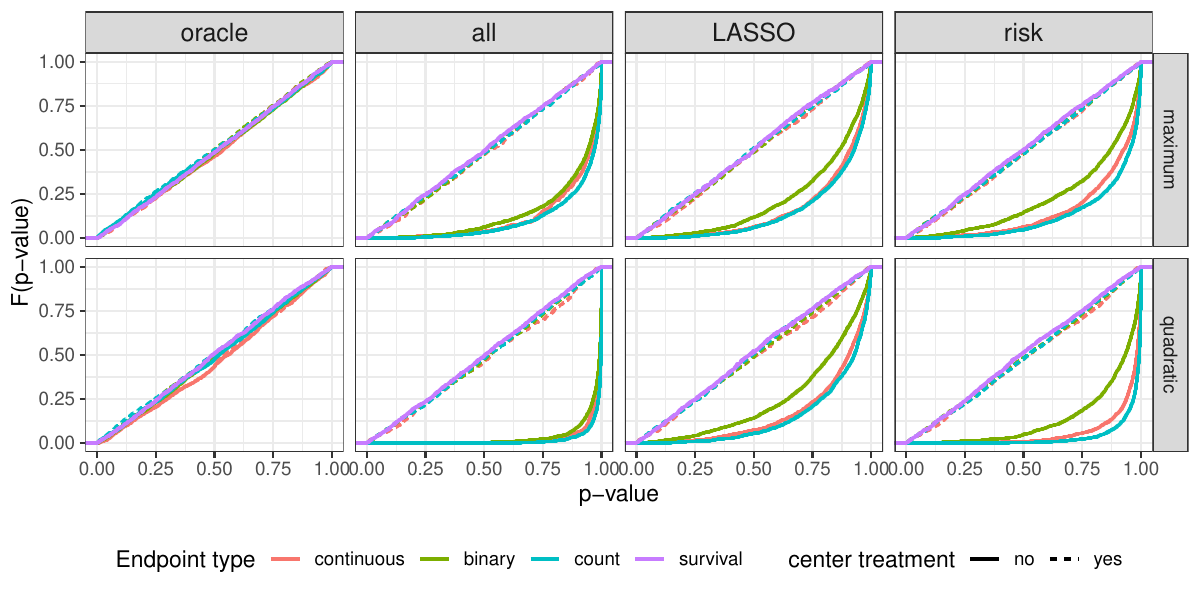}
    \caption{ECDF for $p$-values (F(p-value)) under null hypothesis under different variable selection methods; From left to right, there are 4 different ways for obtaining prognostic effect to derive residual score, oracle, all, lasso, risk, as describe in Section \ref{sec:resid_based}. The top row is the heterogeneity test using maximum statistic, and the bottom row is using quadratic statistic, as described in Section \ref{sec:using-score-residual}. For a uniformly distributed p-values, the ECDF follows a diagonal straight line. For the curves below diagonal line, there are large portion of larger p-values, for curves above the diagonal lines, there are large portion of smaller p-values.}
    \label{fig:pdist_null} 
\end{figure}

% Non-centered not robust under model mis-specification
Firstly, the effect of centering the treatment indicator is investigated. The considerations in Section \ref{sec:resid_based} suggest that the model with centered treatment indication should be more robust when there is model mis-specification on the prognostic variables. The results in Figure \ref{fig:pdist_null} are consistent with these considerations. When there is no model mis-specification, i.e., a model with true prognostic variables (oracle), the ECDFs of $p$-value do not differ much between centered non-centered treatment indicators. All $p$ values are uniformly distributed as expected (the ECDF follows the straight diagonal line). However, in reality, true prognostic variables are not known in advance, but need to be estimated. In the three right columns of Figure \ref{fig:pdist_null}, when the prognostic variables are all variables, selected from the LASSO or predicted risk score, the $p$-values of the model with the non-centered treatment indicator fail to maintain a uniform distribution, while the centered treatment indicator is more robust in such a case. For survival endpoints, the results of the centered versus non-centered treatment indicator are the same, which is due to that the residual score calculation under the centered or non-centered treatment indicator is the same \citep{therneau2015package}. 

% Maximum statistics provide larger power
Both maximum and quadratic test statistics achieve uniform distributions in the situation of homogeneity with centered treatment indicator. This is remarkable as the risk and LASSO based methods use the data twice (first for selecting/estimating prognostic effects and second for performing the global test). 

\begin{figure}[htbp]
    \centering
    \includegraphics[width=0.7\linewidth]{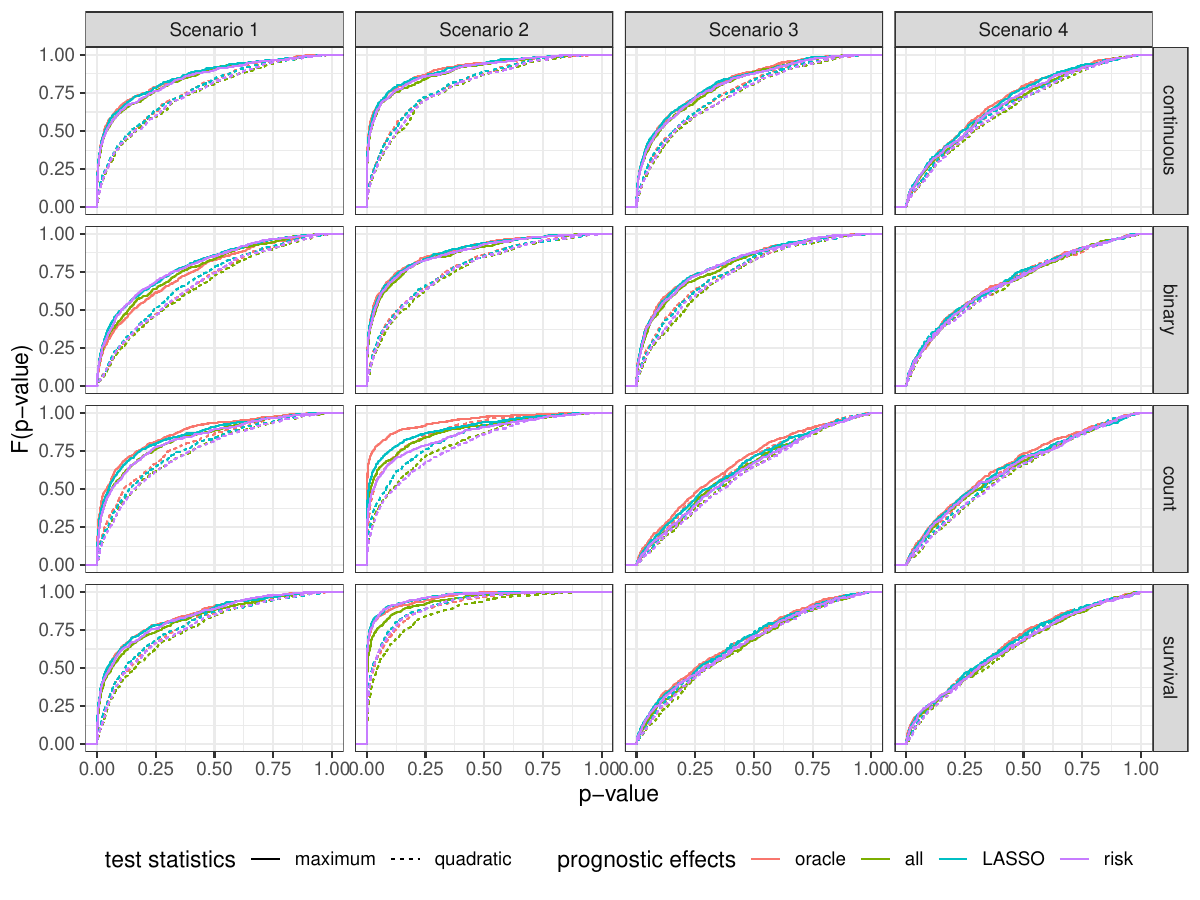}
    \caption{ECDF for $p$-values (F(p-value)) using centered treatment indicators under alternative hypothesis with different endpoint types, data generation scenarios, variable selection type, and global TEH test statistic. The ECDF is based on $p$-value result from 500 replicates. With TEH, a good method would yield a small $p$-value, leading to the ECDF curve above and far away from the diagonal line.}
    \label{fig:pdist_alt} 
\end{figure}

In Figure \ref{fig:pdist_alt}, we demonstrate the distribution of the $p$-values under alternative hypothesis ($\gamma_1/\gamma^*_1 = 2$), i.e. when there exists treatment effect heterogeneity. In such a case, we expect a better method to display a $p$-value distribution shifted towards 0, with an ECDF curve above and far away from the diagonal line. From Figure \ref{fig:pdist_alt}, we can see that in general the maximum test statistic tends to performs better regardless of the types of endpoint or prognostic variables included. This may not be surprising, as in our simulation cases heterogeneity is driven either by one or two variables (so possibly favoring the maximum statistic). In fact the performance of the quadratic statistics for situations where two variables drive heterogeneity (column 3 and 4) seems somewhat closer to the maximum statistic.

% variable selection
In terms of the variables from both Figure \ref{fig:pdist_null} and Figure \ref{fig:pdist_alt}, there is no major difference between using all, LASSO, and risk variables once the treatment indicator is centered. The LASSO method tends to perform closest to the oracle method across the considered scenarios (Figure \ref{fig:pdist_alt}). Note that in our true scenarios there are always two covariates included as prognostic effects and this may explain the slightly better performance of LASSO (which can select some parameter estimates to be exactly 0) compared to the risk based approach with ridge penalty.  

% summary
% In the remaining comparisons against alternative methods we will hence show the residual based methods with centered treatment indicator, maximum test statistic and LASSO-selected variables or risk in the model adjusted as prognostic effects abbreviated as \textit{Residual LASSO} and \textit{Residual Risk}.

\newpage

\subsection{Covariates used for case example}

\begin{table}[ht]
\tiny{
\centering
\begin{tabular}{lll}
  \hline
Names & Description & Categories  \\ 
  \hline
  AGE &  &  \\ 
  ASPRNFL & Baseline Aspirin Therapy & Y, N \\ 
  BASECRP & Log10(High Sensitivity CRP (mg/L)) at Baseline &  \\ 
  ETHNIC &  & HISPANIC OR LATINO, NOT HISPANIC OR LATINO, UNKNOWN \\ 
  GLYCEM & Glycemic Status at Baseline & Diabetic, Normoglycemic, Prediabetic \\ 
  LBEGFG1B & Baseline EGF rate (mL/min) & $<$ 60 mL/min/SA, $>$= 60 to $<$ 90 mL/min/SA, $>$= 90 mL/min/SA \\ 
  LBLOGHDL & Base log value of HDL cholest. (mmol/L) &  \\ 
  LBLOGLDL & Base log value of LDL-C derived (mmol/L) &  \\ 
  LBLOGTRI & Base log value of Triglycerides (mmol/L) &  \\ 
  MHGOUTFL & Medical History of Gout & N, Y \\ 
  QMITGR3 & Time Since Qualifying MI & $<$ 12 months, $>$= 12 months \\
  RACE &  & ASIAN, OTHER, WHITE \\ 
  REGION1 &  & ASIA, CENTRAL EUROPE, LATIN AMERICA \\ 
         &  & NORTH AMERICA, OTHERS, WESTERN EUROPE \\
  SEX &  & F, M \\ 
  SMOKE & Smoking Status at Baseline & Current smoker, Former smoker, Never \\ 
  STATINB & Base median daily dose of statin (mg) & High Dose, Low Dose, Medium Dose, No Dose \\ 
  VSBMIS & Screening BMI (kg/m2) &  \\ 
  VSSTDBMB & Baseline mean sitting diastolic BP (mmHg) &  \\ 
  VSSTSBMB & Baseline mean sitting systolic BP (mmHg) &  \\ 
   \hline
\end{tabular}}
\caption{Naming of baseline variables and used categories}
\label{tbl:variables}
\end{table}

\subsection{Case example individual plot for top select effect modifiers}
To understand how each of the top identified treatment effect modifier contributes to the heterogeneity of treatment effect, the K-M plot for each treatment and treatment effect versus ethnic and baseline mean sitting diatolic BP (mmHg) are displayed in Figure \ref{fig:ind1}. 

For ethnic, there is only slight differences in terms of survival probability between treatment and control at different ethnic levels (Figure \ref{fig:ind1} (a)), which is consistent with the findings of the log hazard ratio of Figure \ref{fig:ind1} (b). Although the UNKNOWN ethnic group has positive log hazard ratio, but due to the relatively small sample size, the variability is large. For Baseline mean sitting diatolic BP, they have negative log hazard ratio (Figure \ref{fig:ind1} (d)) on most values from 60 to 100, which is consistent with the K-M plot from \ref{fig:ind1} (c). 

\begin{figure}[htbp]
    \centering    
    \subfloat[K-M plot on different levels of ethnic.]{%
        \includegraphics[width=0.8\textwidth]{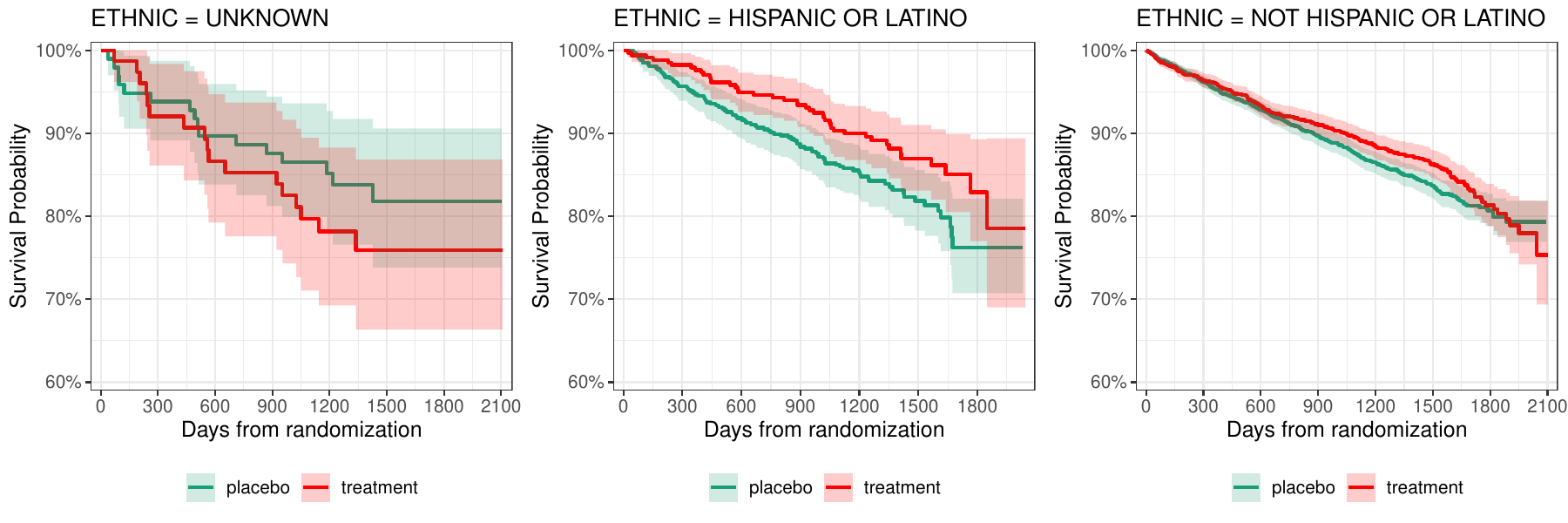}   
    }

    \subfloat[Log hazard ratio on different levels of ethnic.]{%
        \includegraphics[width=0.4\textwidth]{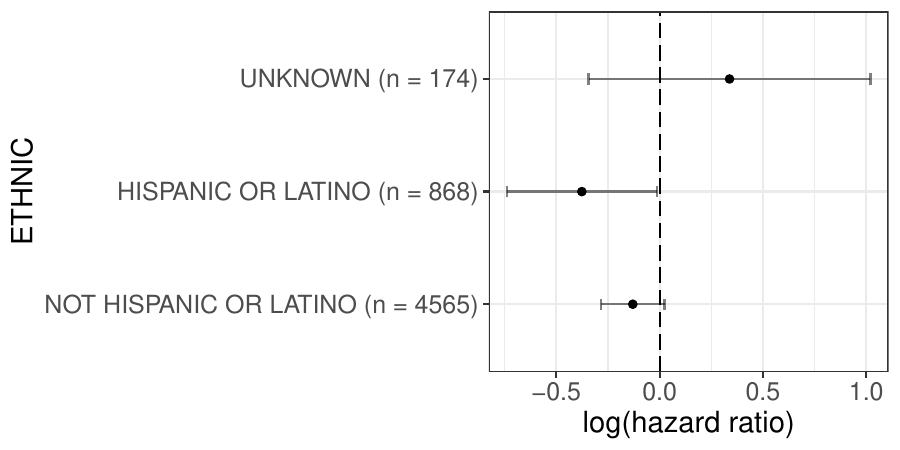} 
    }

    \subfloat[K-M plot on different levels of Baseline mean sitting diatolic BP.]{%
        \includegraphics[width=0.5\textwidth]{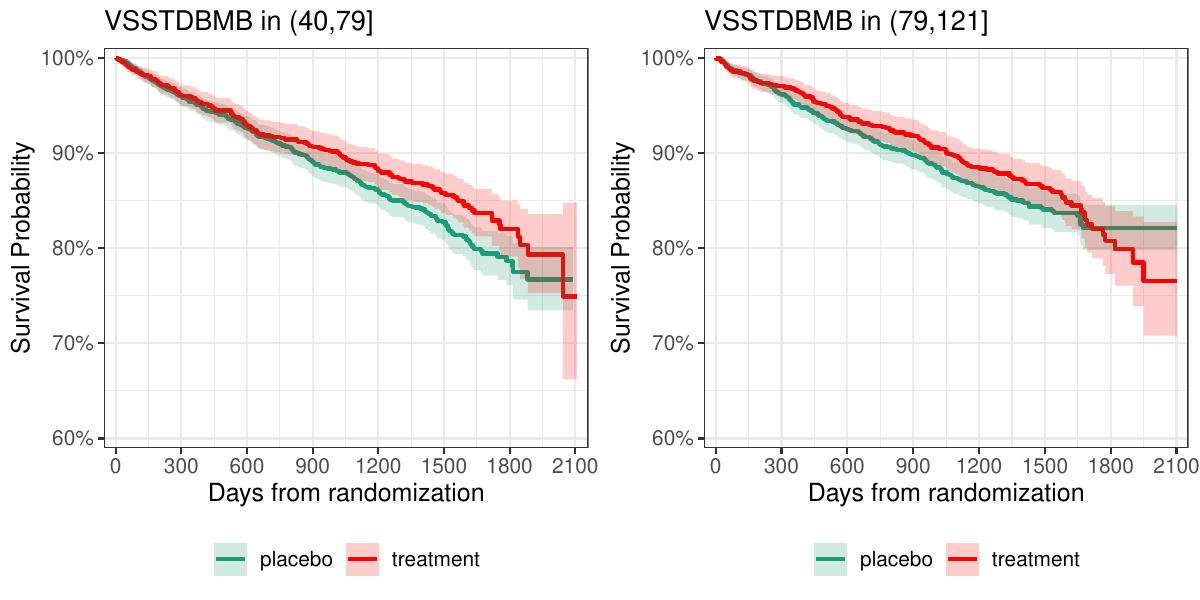}
    }
    \subfloat[Log hazard ratio on different levels of Baseline mean sitting diatolic BP.]{%
        \includegraphics[width=0.3\textwidth]{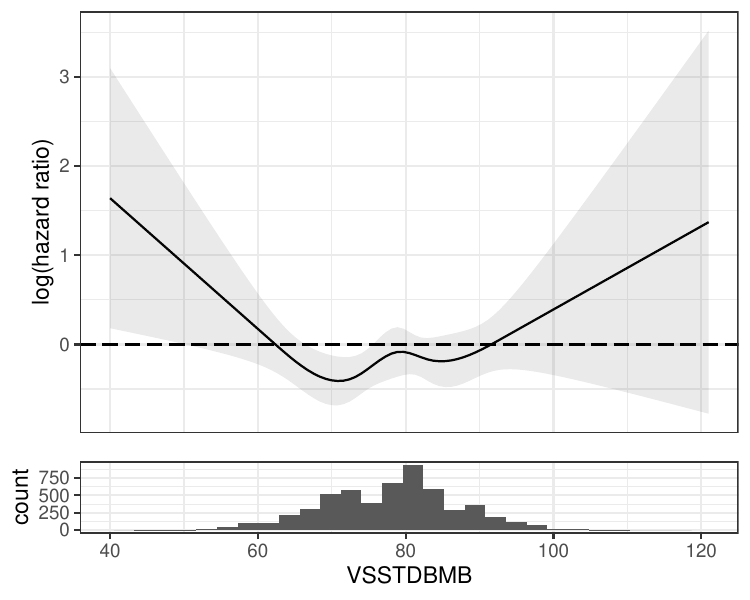}
    }
    
    \caption{K-M curve/treatment effect on ethnic and baseline mean sitting diatolic BP. For (b), the point and confidence interval are calculated based on Cox model fitted separately on each ethic level. For (d), the curve and confidence interval is calculated based on spline fitting of response versus identified modifier.}
    \label{fig:ind1}
\end{figure}

Based on the important effect modifiers and their relationship with endpoints, we see there is no strong evidence for heterogeneity, due to the large variability. These findings are consistent with the result from treatment effect heterogeneity test.

\subsection{Extra example}
To illustrate the methods in this paper we will use synthetic, simulated data that mimic data from a cardiovascular outcome study, which assessed three investigational doses versus placebo with cardiovascular outcomes as the primary endpoint. The data contains 3300 patients on the placebo group and 2131, 2240, 2195 patients on each of the investigational treatment arms, corresponding to doses of 50, 150 and 300 mg. For the purpose here we will consider an outcome variable that was an exploratory endpoint in the study. The investigational drug is targeted to lower inflammation, so it is of interest to confirm the effect of the treatment on an inflammation marker at 3 months post baseline. For this purpose the used outcome variable will be the base 10 logarithm high-sensitivity C-reactive protein (hs-CRP) change from baseline.  

\begin{figure}[htbp]
    \centering
    \includegraphics[width=0.4\linewidth]{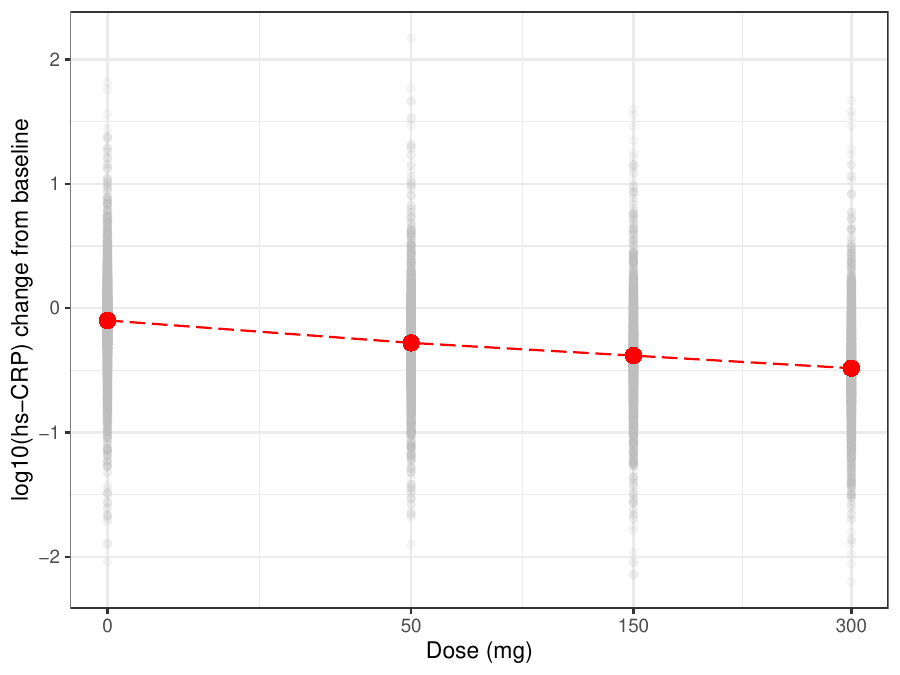}
    \caption{Logarithm of hs-CRP change from baseline on dose (mg). The gray points are the observed data and the red points are the average for each dose level.}
    \label{fig:crpraw} 
\end{figure}

In Figure \ref{fig:crpraw} one can see the observed data overlaid with corresponding dose-specific means on a square-root scale for the dose. To adequately describe these dose-response data we will hence utilize the following dose-response model

\begin{equation}
  \label{eq:dr_mod}
    y_{i}=\beta_0+\mathbf{x}_i'\boldsymbol{\beta} + \delta (\sqrt{d}_i-d_m) +\epsilon_i,\;\mathrm{with}\;\epsilon_i \sim N(0,\sigma^2), \;i=1,\ldots,N,
\end{equation}
here $d_m = \frac{1}{N}\sum_{i=1}^N \sqrt{d}_i$. The treatment effect (difference versus placebo) for a dose $d$ in this model is hence given by $\delta\sqrt{d}$. The reason for centering the square root of the dose (similar as in Section \ref{sec:methodology}) is that the intercept in the model will correspond to the overall mean, rather than the placebo group mean. This condenses the information on all group means directly in the single parameter $\delta$. To explore treatment effect heterogeneity 19 potential effect modifiers will be utilized, described in detail in Table \ref{tbl:variables} in the Appendix. These variables were selected based on the subgroup variables pre-defined for subgroup analyses for the primary endpoint. The vector $\mathbf{x}$ is hence an 19-dimensional vector for each patient, containing the potential effect modifiers. 

We fitted the model \eqref{eq:dr_mod} in two variations, V1: including all 19 covariates as well as treatment in the model and fitting the model by maximum likelihood. V2: first fitting a penalized regression without treatment indicator, using a ridge penalty and choosing the penalty parameter according to 10-fold cross-validation. Based on the penalized model then a model is fitted adjusting for intercept, dose (as in equation \eqref{eq:dr_mod}) and the resulting risk score as prognostic effect. For both versions, the score residual for treatment indicator is extracted. In addition we will utilize the MOB methodology with a non-centered treatment indicator adjusting in the same way as described in V1 and V2, and extract the $p$-value for the fluctuation test in the root node.

The resulting model fit on the dose-response scale for V1 can be observed in Figure \ref{fig:crpraw}, showing that the data are adequately described by the square-root of dose in the observed dose-range.

When using permutation independence test as a global test on score residuals for the slope of the square-root of the dose versus all covariates, the test results in $p$-values of 0.011 (V1) and 0.0053 (V2) for the quadratic and 7.8e-05 (V1), 6.60e-05 (V2) for the maximum test statistic. On the surprise scale (-log2(p)) these correspond to 6.6, 7.6, 13.6, 13.9, which according to the scale in \cite{sechidis2024watch} corresponds to strong/very strong evidence against the null-hypothesis of treatment effect homogeneity. The result is quite consistent compared to that of MOB which gives a $p$-values of 0.012 (surprise value 6.4) using risk as covariate (V1) and 4.31e-06 (surprise value 17.8) using all covariates (V2). 

Variable importance measures have been calculated by fitting a conditional random forest \cite{hothorn2006unbiased} to the score residuals (based on the model fit \eqref{eq:dr_mod}) and calculating the permutation importance of the underlying variables (see Figure \ref{fig:vi}). The variable importance plot from both versions are quite consistent, both choosing screening BMI, Log10(hs-CRP (mg/L)) at baseline, baseline EGF rate (mL/min), and baseline log value of Triglycerides (mmol/L) as the top 4 important variables. 

\begin{figure}[htbp]
    \centering
    \subfloat[Variable importance for V1]{%
        \centering
        \includegraphics[width=0.35\textwidth]{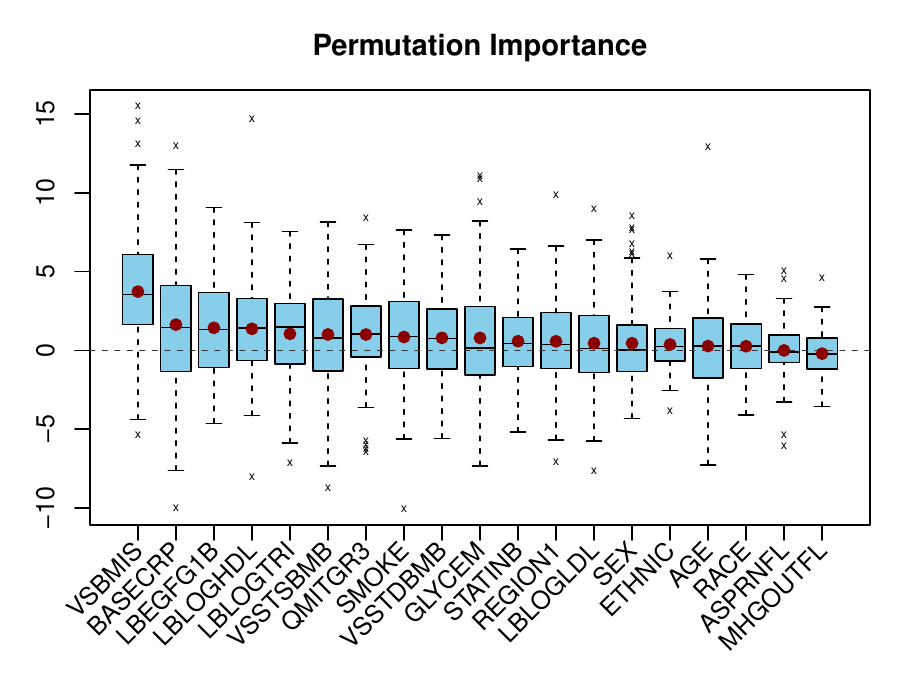}
    }
    \subfloat[Variable importance for V2]{%
        \includegraphics[width=0.35\textwidth]{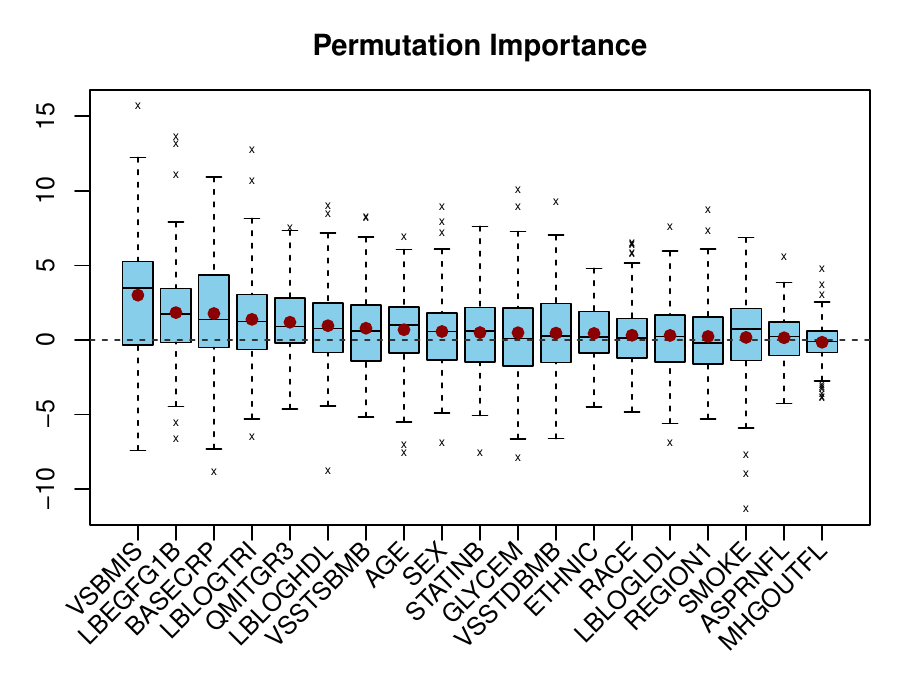}
    }
    \caption{Variable importance for treatment effect modifiers for two versions of models. For each boxplot, each point is the permutation importance for each tree based on conditional random forest with 100 tress.}
    \label{fig:vi}
\end{figure}

To better understand how each of the top identified treatment effect modifier contributes to the heterogeneity of treatment effect, the estimated outcome on each of the doses and treatment effect versus screening BMI and baseline CRP are displayed in Figure \ref{fig:ind1}. 

For screening BMI, the gap between placebo outcome versus outcome on other dosages changes as BMI increases (Figure \ref{fig:ind1} (a)), combined with the treatment effect plot Figure \ref{fig:ind1} (b), we can see that the treatment effect (on each dosage versus placebo) in general slightly decreases, then increases as BMI increases, in particular for the higher dose groups. For the log10(hs-CRP) at baseline, the treatment effect decreases first, then increases (Figure \ref{fig:ind1} (d)). 

\begin{figure}[htbp]
    \centering
    \subfloat[Outcome versus screening BMI. ]{%
        \includegraphics[width=0.35\textwidth]{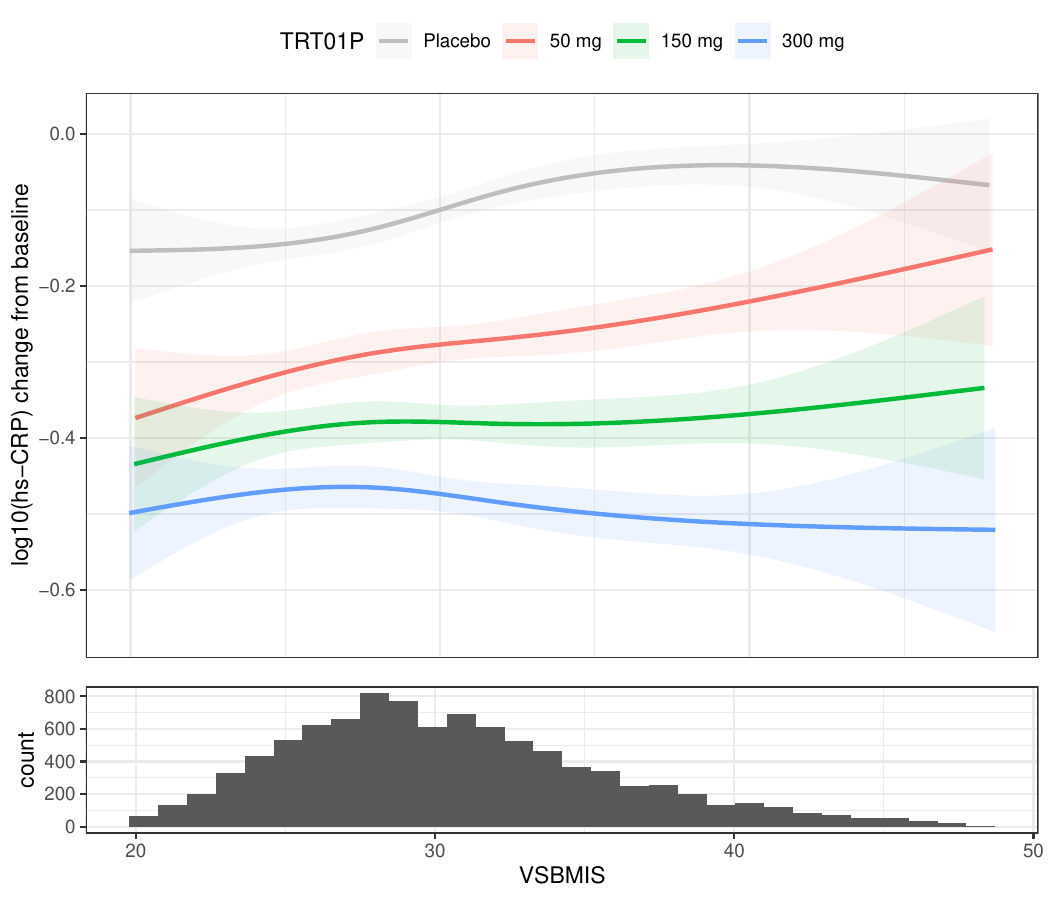}
    }
    \subfloat[Treatment effect versus screening BMI.]{%
        \includegraphics[width=0.35\textwidth]{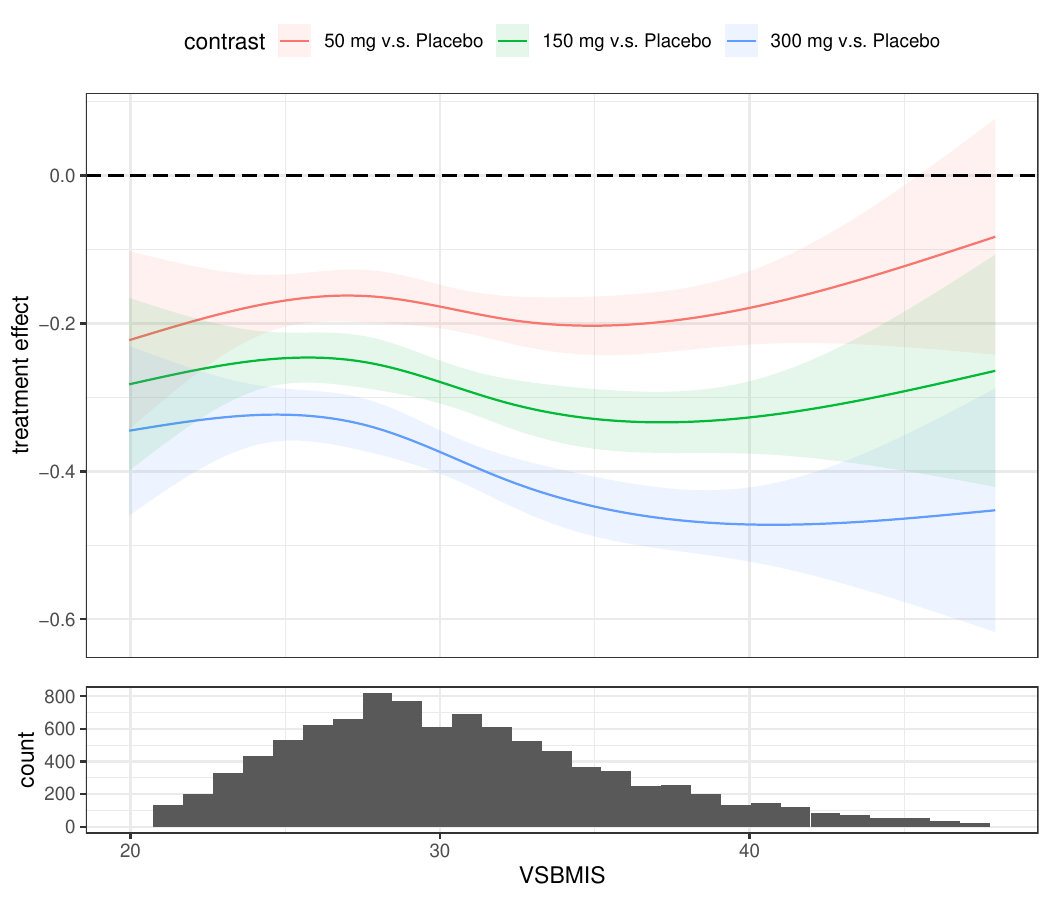}
    }

    \subfloat[Outcome versus Log10(hs-CRP) at baseline.]{%
        \includegraphics[width=0.35\textwidth]{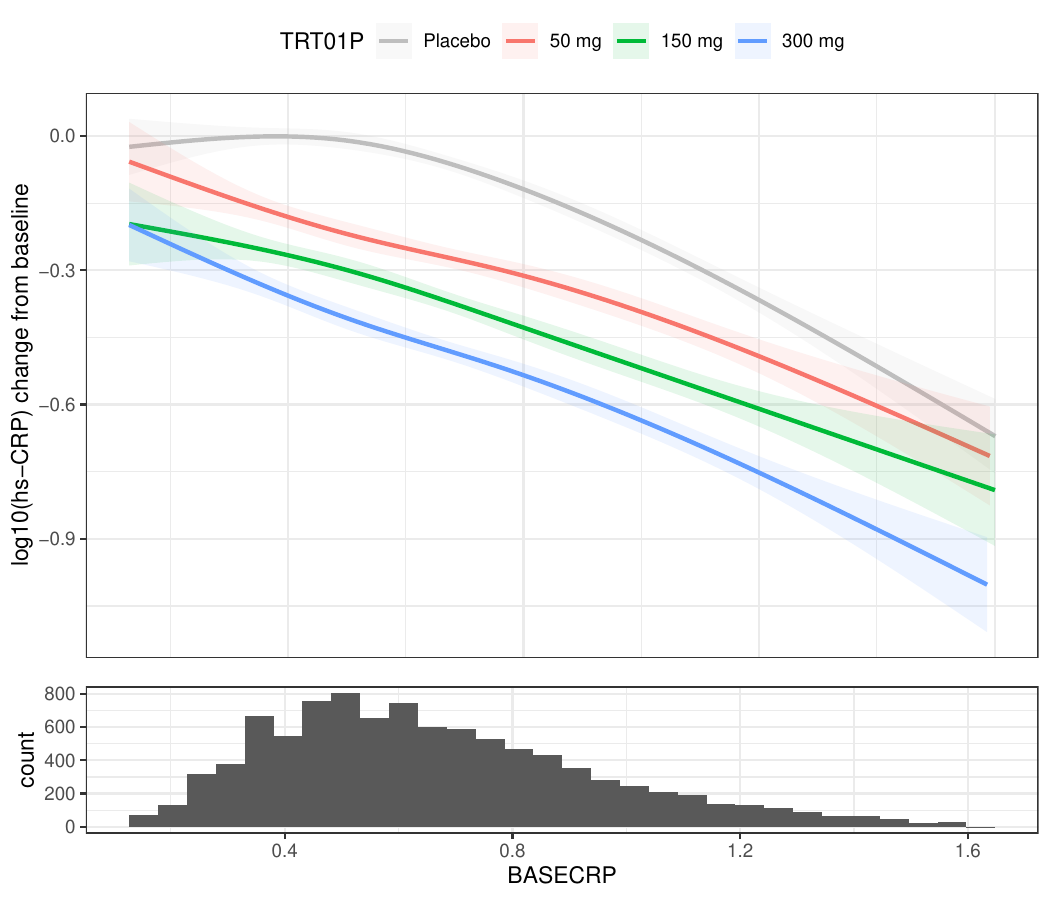}
    }
    \subfloat[Treatment effect versus Log10(hs-CRP) at baseline.]{%
        \includegraphics[width=0.35\textwidth]{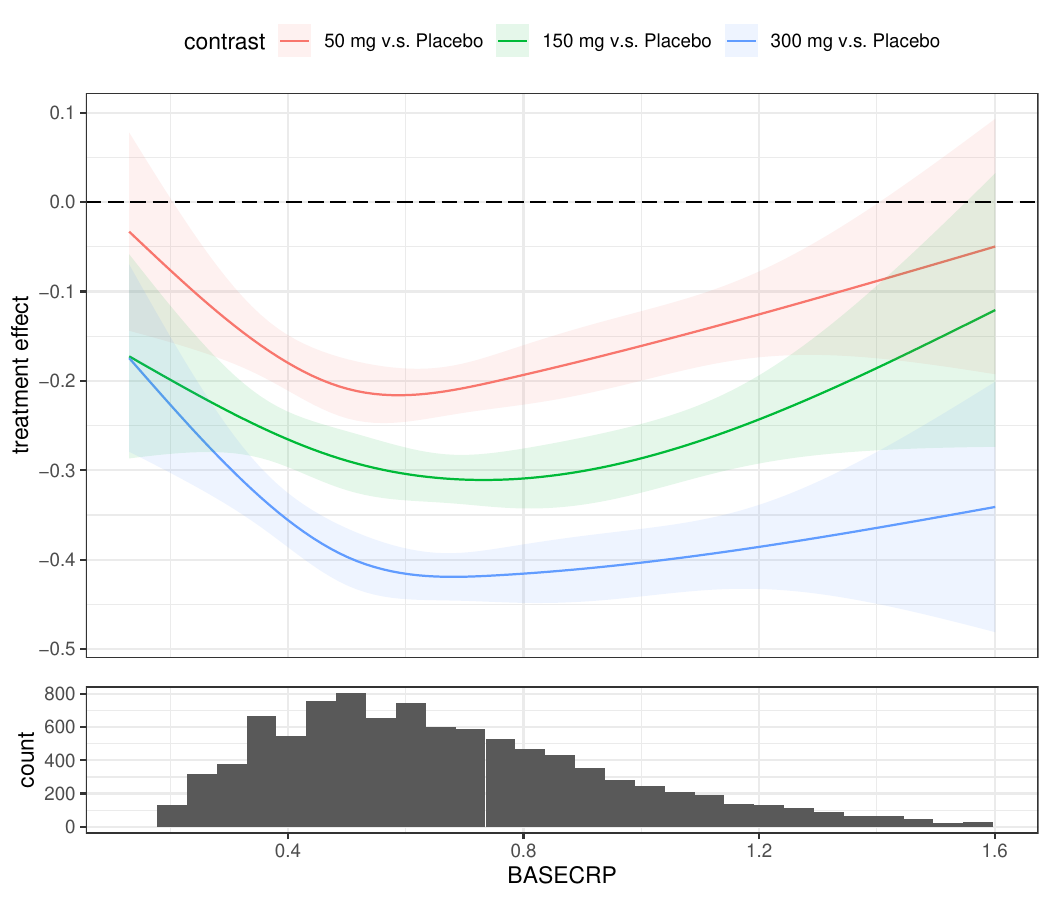}
    }
    
    \caption{Response/treatment effect on screening BMI, Log10(hs-CRP (mg/L)) at baseline. The curve and confidence interval is calculated based on spline fitting of response versus identified modifier.}
    \label{fig:ind1}
\end{figure}

Based on the observed data there is strong evidence against a homogeneous treatment effect. This is not entirely surprising, as in this case the study was sample-sized for a primary time-to-event endpoint and the power for detecting differences in this biomarker endpoint are much larger. In fact the sample size is around 100-fold larger than needed for showing a treatment effect in hs-CRP. The sample size is hence considerably larger than one would commonly encounter in an analysis of treatment effect heterogeneity, which often targets the primary endpoint in a randomized clinical trial (and clinical trials are generally sample-sized for the primary endpoint). In this case the practical relevance of the magnitude of treatment effect changes across the identified variables hence needs to be assessed critically, as there may be high evidence against homegeneity, but the observed heterogeneity may however not need to be practically relevant.

\bibliographystyle{unsrtnat}
\bibliography{bibl.bib}

\end{document}